\documentclass[1p,fleqn,times]{elsarticle}

\usepackage{hyperref}

\usepackage{geometry} % Required for adjusting page dimensions

\usepackage{natbib}
\usepackage{amsmath,amssymb}
\usepackage{mathrsfs}
\usepackage{physics}
\usepackage{xcolor}
\usepackage{mathtools}
\usepackage{graphicx}
\usepackage{rotating}
\usepackage{blindtext}
\usepackage{multirow}
\usepackage{comment}
\usepackage{subcaption}
\usepackage{ulem}

\newcommand{\Vbar}{\overline{\mathbf{V}}^e}%{\overline{V}^e}

\newcommand{\dev}[1]{\mbox{dev}\left({#1}\right)}
\newcommand{\Lagrangian}{\mathbf{X}}
\newcommand{\normal}{\mathbf{\hat{n}}}

\graphicspath{{}}

\makeatletter
\def\ps@pprintTitle{%
  \let\@oddhead\@empty
  \let\@evenhead\@empty
  \def\@oddfoot{}%
  \let\@evenfoot\@oddfoot}
\makeatletter
\def\ps@pprintTitle{%
  \let\@oddhead\@empty
  \let\@evenhead\@empty
  \def\@oddfoot{}%
  \let\@evenfoot\@oddfoot}
\makeatother
\usepackage{fancyhdr}

\fancypagestyle{pprintTitle}{
\lhead{}\chead{}\rhead{}\lfoot{\textcopyright \small ~British Crown Owned Copyright 2022/AWE}\cfoot{\thepage}\rfoot{}
}

\begin{document}

\begin{frontmatter}

\title{A multi-physics method for fracture and fragmentation at high strain-rates}

\author[mymainaddress]{Tim Wallis \corref{mycorrespondingauthor}}
\cortext[mycorrespondingauthor]{Corresponding author}
\ead{tnmw2@cam.ac.uk}

\author[mysecondaryaddress]{Philip T. Barton}
\author[mymainaddress]{Nikolaos Nikiforakis}

\address[mymainaddress]{Department of Physics, University of Cambridge, Cavendish Laboratory, JJ Thomson Avenue, CB3 0HE, UK}
\address[mysecondaryaddress]{AWE Aldermaston, Reading, Berkshire, RG7 4PR, UK}

\begin{abstract}
This work outlines a diffuse interface method for the study of fracture and fragmentation in ductile metals at high strain-rates in Eulerian finite volume simulations. The work is based on an existing diffuse interface method capable of simulating a broad range of different multi-physics applications, including multi-material interaction, damage and void opening. The work at hand extends this method with a technique to model realistic material inhomogeneities, and examines the performance of the method on a selection of challenging problems. Material inhomogeneities are included by evolving a scalar field that perturbs a material's plastic yield stress. This perturbation results in non-uniform fragments with a measurable statistical distribution, allowing for underlying defects in a material to be modelled. As the underlying numerical scheme is three dimensional, parallelisable and multi-physics-capable, the scheme can be tested on a range of strenuous problems. These problems especially include a three-dimensional explosively driven fracture study, with an explicitly resolved condensed phase explosive. The new scheme compares well with both experiment and previous numerical studies.
\textcopyright ~British Crown Owned Copyright 2022/AWE
\end{abstract}


\end{frontmatter}

\pagestyle{pprintTitle}
\thispagestyle{pprintTitle}

\section{Introduction}

\begin{comment}
\begin{itemize}
 \item \sout{Abstract}
 \item \sout{Introduction}
 \item Governing Theory
 \subitem \sout{Evolution Equations}
 \subitem Closure Models
 \subitem Plasticity
 \subitem Damage
 \subitem Lagragian field
 \item Numerical Method
 \subitem Damage (and Plasticity?)
 \subitem Lagrangian field
 \item Validation
 \subitem 1D Spallaiton
 \subitem Expanding Disk
 \subitem Expanding Ring
 \subitem Explosive shell
 \item \sout{Conclusions}
 \item \sout{Material Parameter appendix}
\end{itemize}
\end{comment}

The dynamic fracture and fragmentation of materials is a well-studied yet highly challenging field for numerical simulation. Fracture simulations play an important role in many industries, including aerospace and automotive safety, mining simulation, geological fracture analysis, and explosive-structure interaction. Physically, damage reduces the load carrying ability of a solid material until, at a critical level, the material loses all strength and fractures. Various branches of physics deal with damage on different scales, depending on the application and type of damage at hand. At the most fundamental level, damage starts as the dislocation of bonds at the atomistic scale. Many authors are concerned with the study of such damage (such as \citet{AtomisticFracture}), and typically employ molecular dynamics simulations. Above this, fracture is present at the micro-scale as the nucleation, growth, and coalescence of micro-voids in the crystal structure of the material \cite{MicroscopicFracture}. Finally, at the macroscopic level, damage is observed as the visible fractures within a material, and is intimately linked to plastic effects. This work will focus on the continuum/macro-scale formation and evolution of damage, rather than attempting to model the microscopic dynamics. 

Difficulties arise for numerical simulations of fracture in the generation of the complex new material interfaces that appear at unknown locations as a result of propagating, branching cracks. These cracks must be created, grow and combine in a consistent manner. As such, fracture problems can strenuously test a numerical method. At present, there are numerous different approaches to fracture that are currently used in the literature. Lagrangian methods have typically been used to model solid dynamics problems, often with finite element methods, and have been applied to fracture in several studies \cite{BeckerRingFracture, MeulbroekRingFracture, ZhouMolinariFEFracture, CamachoFEFracture}. However, a common approach in these methods is to initiate fracture by removing damaged elements to form cracks. Although this approach does yield the effects of fracture, it is also highly non-conservative as mass, momentum and energy are removed from the domain when the element is removed. An alternative approach to element removal is the cohesive zone method \cite{OrtizCohesiveZone, CirakCohesiveZone}, where cracks can be generated along element boundaries. This approach has better conservation properties but is highly mesh-dependent. Lagrangian methods can also struggle with the severe mesh distortion present in the high-strain rate, high-deformation problems examined here. These issues have led to the development of mesh-free approaches \cite{LiExplosiveShell, RabczukMeshFree}, peridynamics \cite{SillingPeridynamics, SillingAskariPeridynamics, RenPeridynamics} and smoothed particle hydrodynamics methods \cite{OwenExpandingDisk, OwenExplosiveShell, RabczukSPH}, which in turn have considered fracture. There also exist many Eulerian approaches to fracture, which do not suffer from re-meshing issues with large distortion. Several Eulerian approaches employ level-sets to track the material boundaries formed by fracture \cite{BartonLevelSetDamage, UdaykumarDamage, ParticleLevelSetDamage}. However, these methods suffer from the same `element erosion' problem as Lagrangian finite element methods, albeit in a different guise, when damaged cells are given a negative level-set value and `removed' from the material.

Diffuse interface methods for fracture in Eulerian schemes also exist \cite{DumbserDamage, NdanouDiffuseFracture, WallisFluxEnriched}. These schemes avoid the material erosion problem by allowing material interfaces to be represented as a continuous scalar field which can continuously open to form a crack. This then avoids many of the conservation issues of deleting damaged cells. Moreover, as the dynamics of material interfaces are already hard-coded into the system of equations for a diffuse interface method, no additional techniques are required to track the growth and possible coalescence of cracks. This removes the need for explicit modelling of crack tip speed, geometry or connectivity. This results in a robust, straightforward scheme that can easily be extended to multiple spatial dimensions with arbitrarily complex fracture.

This work extends the diffuse interface fracture method of \citet{WallisFluxEnriched} to model realistic material inhomogeneity. The underlying method is a multi-material diffuse interface scheme based on the \citet{Allaire} multi-fluid model. The model has been subsequently extended a number of times to include a range of different physics: first by \citet{Barton2019} to encompass solid dynamics, then by \citet{WallisMultiPhysics} to include reactive fluid mixtures using the work of \citet{MiNi16}, then by \citet{WallisFluxEnriched} to include damage, fracture and void opening, and most recently by \citet{WallisRigidBody} to include rigid bodies. All these developments allow the model to study a range of real-world problems, thanks to its broad applicability.

The method outlined so far \cite{WallisFluxEnriched} treats materials as purely homogeneous and isotropic. Damage will therefore grow uniformly inside such materials as a result. However, this is not observed in experiment. Any real-world material will contain defects and inhomogeneities in its structure. This can lead to damage localisation around flaws and strengthening in other regions, ultimately leading to a more anisotropic fragment distribution. So far, the micro-structure of the material has been neglected in the development of the model. Unfortunately, these inhomogeneities are precisely a result of the micro-structure of the material, and damage localisation is inherently a not a continuum-scale process. Therefore the challenge presents itself: how can material inhomogeneities be accounted for in a continuum damage model?

This work achieves this goal by perturbing the constitutive material models, using an approach similar to \citet{VitaliTransportDiffusion}. This method, initially developed for arbitrary Lagrangian Eulerian (ALE) methods, includes a randomly varying scalar field $\varphi$ to model the underlying defects in any material. This scalar field can be initialised to any desired distribution, or even empirically determined for a given material. This perturbation field then modulates the plastic yield stress of the material, acting to locally raise and lower the strength of the material, mimicking the effect of inhomogeneities. 

This paper will proceed as follows. Firstly, the background of the \citet{WallisFluxEnriched} fracture model will be briefly outlined, followed by the description of the material inhomogeneities for damage perturbation. Then the numerical methods used will be outlined, followed by validation via a number of fracture problems, and finally conclusions will be drawn. 

\section{Governing Theory}

\subsection{Evolution Equations}

The base system, outlined by \citet{WallisFluxEnriched}, is an Allaire-type \cite{Allaire} multi-material diffuse interface system, capable of accounting for an arbitrary mixture of both fluids and elastoplastic solids. The state of any material $l$ is characterised by its phasic density $\rho_{(l)}$, volume fraction $\phi_{(l)}$, symmetric left unimodular stretch tensor $\Vbar$, velocity vector $\mathbf{u}$, internal energy $\mathscr{E}$, and history variable vector $\boldsymbol{\alpha}_{(l)}$.

The left stretch tensor $\mathbf{V}^e$ is related to the deformation tensor $\mathbf{F}^e$ by the polar decomposition:
\begin{equation}
 \mathbf{F}^e = \mathbf{V}^e\mathbf{R}^e,
\end{equation}
after which it is normalised to obtain $\Vbar$:
\begin{equation}
 \Vbar = \det\left(\mathbf{V}^e\right)^{-1/3}\mathbf{V}^e.
\end{equation}

The simplifying assumption is made that materials in a mixture are in mechanical and thermal equilibrium, resulting in a reduced equation system where materials share the same momentum, energy and deformation equations. This would normally limit the model to modelling solely `stick' boundary conditions, however the flux-modifiers developed by \citet{WallisFluxEnriched} allow the model to incorporate slip and void-opening conditions as well.

For $l=1,\ldots, N$ materials:
\begin{align}
\frac{\partial \phi_{(l)}}{\partial t} + \frac{\partial \phi_{(l)} u_k}{\partial x_k} &= \phi_{(l)}\frac{\partial u_k}{\partial x_k}\\
\frac{\partial \phi_{(l)}\rho_{(l)}}{\partial t} + \frac{\partial \phi_{(l)}\rho_{(l)} u_k}{\partial x_k} &= 0 \\
\frac{\partial \phi_{(l)}\rho_{(l)}\boldsymbol{\alpha}_{(l)}}{\partial t} + \frac{\partial \phi_{(l)}\rho_{(l)} \boldsymbol{\alpha}_{(l)} u_k}{\partial x_k} &= \phi_{(l)}\rho_{(l)}\dot{\boldsymbol{\alpha}}_{(l)} \\
\frac{\partial \rho u_i }{\partial t} + \frac{\partial (\rho u_i u_k -\sigma_{ik}) }{\partial x_k} &= 0\\
\frac{\partial \rho E }{\partial t} + \frac{\partial (\rho E u_k - u_i \sigma_{ik}) }{\partial x_k} &= 0\\
\frac{\partial \Vbar_{ij} }{\partial t} + \frac{\partial \left( \Vbar_{ij}  u_k - \Vbar_{kj} u_i \right) }{\partial x_k} &= \frac{2}{3}\Vbar_{ij}\frac{\partial u_k}{\partial x_k} - u_i\beta_j - \boldsymbol\Phi_{ij}  \ ,
\end{align}
here $E=\mathscr{E}+|\mathbf{u}|^2/2$ denotes the specific total energy, $\boldsymbol\sigma$ denotes the Cauchy stress tensor, $\beta_j=\partial\Vbar_{kj}/\partial x_k$, $\boldsymbol\Phi$ represents the contribution from plastic effects.

The plastic source term $\boldsymbol\Phi$ limits the possible range of elastic deformation and initiates the growth of plastic deformation. This work follows the method of convex potentials, defining this source term as 
\begin{equation}
 \boldsymbol\Phi = \chi\pdv{\varphi(\boldsymbol\sigma)}{\boldsymbol\sigma}\Vbar \ ,
\end{equation}
where $\chi$ is a plastic multiplier and $\varphi$ is a convex potential, both of which are closure models. This source term is outlined in previous works \cite{Barton2019,WallisMultiPhysics,WallisFluxEnriched}.

The requirement to model realistic materials will necessarily require material closure models introduce dependencies on material history variables, $\boldsymbol{\alpha}$. In this work, these will especially include the effective equivalent plastic strain $\varepsilon_{p,(l)}$ and the scalar damage parameter $D_{(l)}$ for metals, and the reaction progress variable $\lambda$ for reactive fluid mixtures, following the methods outlined in \citet{WallisMultiPhysics} and \citet{WallisFluxEnriched}. Additional evolution equations are required to advect and evolve these variables as time progresses.

As well as the thermodynamic variables outlined above, certain `mechanical' variables are also required for the application to fracture:
\begin{align}
\frac{\partial \nu }{\partial t} + \frac{\partial \nu u_k}{\partial x_k} &= \nu \frac{\partial u_k}{\partial x_k}  \\
\frac{\partial r}{\partial t} + (\mathbf{v}_r)_k\frac{\partial r }{\partial x_k} &= 0 \\
\pdv{X_i}{t} + \pdv{X_iu_k}{x_k} &= 0 \ ,
\end{align}
here, $\nu$ is the void volume fraction, $r$ is the rigid body volume fraction and $\Lagrangian$ are the Lagrangian material coordinates.
The additional volume fraction fields $\nu$ and $r$ are required to facilitate the void-opening, fracture, and rigid body methods outlined in \citet{WallisFluxEnriched} and \citet{WallisRigidBody}. The void volume fraction is evolved in exactly the same way as material volume fractions. On the other hand, the rigid body volume fraction is evolved by its own velocity field $\mathbf{v}_r$. This velocity field can be any required function of space and time, but it does not mutually interact with the flow at large; it is simply imposed on the rigid body. Finally, the method for modelling material inhomogeneities requires that the material coordinates are tracked over the course of the simulation. This will be further explained in Section \ref{sec:PerturbationField}.

\subsubsection{Thermodynamics}
\label{sec:thermodynamics}

The thermodynamics of undamaged materials have been presented before at length \cite{Barton2019,WallisMultiPhysics,WallisFluxEnriched,WallisRigidBody}, so this section will focus on outlining the thermodynamics of damageable materials and the background of the damage model. 

It will be assumed that the internal energy $\mathscr{E}$ for each material is defined by an equation-of-state that conforms to the general form:
\begin{equation}\label{eq_eos_gen}
\mathscr{E}_{(l)}\left(\rho_{(l)},T_{(l)},\dev{\mathbf{H}^e},\boldsymbol{\alpha}_{(l)}\right) = \mathscr{E}_{(l)}^c\left(\rho_{(l)},\boldsymbol{\alpha}_{(l)}\right)+ \mathscr{E}_{(l)}^t\left(\rho_{(l)},T_{(l)}\right) + \mathscr{E}_{(l)}^s\left(\rho_{(l)},\dev{\mathbf{H}^e},\boldsymbol{\alpha}_{(l)}\right) \ ,
\end{equation}
where
\begin{equation}
\dev{\mathbf{H}^e} = \ln\left(\Vbar\right) \ 
\end{equation}
is the deviatoric Hencky strain tensor and $T$ is the temperature. The three terms on the right hand side are respectively the contribution due to volumetric strain (cold-compression and dilation), $\mathscr{E}_{(l)}^c$, the contribution due to temperature deviations, $\mathscr{E}_{(l)}^t$, and the contribution due to shear strain $\mathscr{E}_{(l)}^s$. The volumetric and shear strain energies will generally be provided by the specific closure model for each material. The thermal energy is given by:
\begin{eqnarray}
\mathscr{E}_{(l)}^t(\rho_{(l)},T)&=& C_{(l)}^{\text{V}}\left(T-T_{(l)}^0\theta_{(l)}^D\left(\rho_{(l)}\right)\right) \ ,
\end{eqnarray}
where $C_{(l)}^{\text{V}}$ is the heat capacity, $T_{(l)}^0$ is a reference temperature, and $\theta_{(l)}^{\text{D}}(\rho_{(l)})$ is the non-dimensional Debye temperature, a closure model.
The Debye temperature is related to the Gr\"uneisen function, $\Gamma(\rho_{(l)})$, via
\begin{equation}
\Gamma_{(l)}(\rho_{(l)}) = \frac{\partial \ln\theta_{(l)}^{\text{D}}(\rho_{(l)})}{\partial \ln(1/\rho_{(l)})} = \frac{\rho_{(l)}}{\theta_{(l)}^{\text{D}}(\rho_{(l)})}\frac{\partial \theta_{(l)}^{\text{D}}(\rho_{(l)})}{\partial\rho_{(l)}} \ .
\end{equation}
The specific form of the Gr\"uneisen function for each material derived from the Debye temperature.
It will also be assumed that the shear internal energy is given by the form:
\begin{equation} \label{eq:shearEnergy}
\mathscr{E}_{(l)}^s(\rho_{(l)},\dev{\mathbf{H}^e},\alpha_{(l)}) =  \frac{G_{(l)}\left(\rho_{(l)},\alpha_{(l)}\right)}{\rho_{(l)}}  \mathcal{J}^2\left(\dev{\mathbf{H}^e}\right) \ ,
\end{equation}
where $G_{(l)}(\rho_{(l)},\alpha_{(l)})$ is the shear modulus, and
\begin{equation}
\mathcal{J}^2(\dev{\mathbf{H}^e}) = \tr\left(\dev{\mathbf{H}^e}\dev{\mathbf{H}^e}^{\text{T}}\right)
\end{equation}
is the second invariant of shear strain. In this case, the specific form of $G$ is a closure model. For each component, the Cauchy stress, $\boldsymbol\sigma$, and pressure, $p$, are inferred from the second law of thermodynamics and classical arguments for irreversible elastic deformations:
\begin{eqnarray}
\boldsymbol\sigma_{(l)} &=& p_{(l)}\mathbf{I} + \dev{\boldsymbol\sigma_{(l)}} \\
p_{(l)} &=& \rho^{2}_{(l)} \frac{\partial\mathscr{E}_{(l)}}{\partial \rho_{(l)}} \label{eq:p_energy_derivative}\\ 
\dev{\boldsymbol\sigma_{(l)}} &=& 2G_{(l)}\dev{\mathbf{H}^e} \ .
\end{eqnarray}
Although it might appear that the model describes solid materials, fluids can be considered a special case where the shear modulus is zero, resulting in a spherical stress tensor and no shear energy contribution. This formulation lends itself well to diffuse interface modelling where different phases that share the same underlying model can combine consistently in mixture regions.

Further details of the thermodynamics of the base system are given in previous works \cite{Barton2019,WallisMultiPhysics,WallisFluxEnriched,WallisRigidBody}.

A common set of methods used for the study of macro-scale damage formation is continuum damage mechanics (CDM). CDM is a thermodynamically consistent way of modelling the effect of damage at the continuum scale, where thermodynamic forces are derived using the second law in order to advect and evolve damage in a given material \cite{MurakamiCDM}.

This work follows \citet{Bonora} in employing a scalar damage model. In this model, the damage variable, $D$, is defined as the fractional area of micro-voids on a plane intersecting a representative volume element, as shown in Figure \ref{fig:DamageVolumeElement}. From this definition, damage is taken to be:

\begin{figure}
 \centering
 \includegraphics[width=0.5\textwidth]{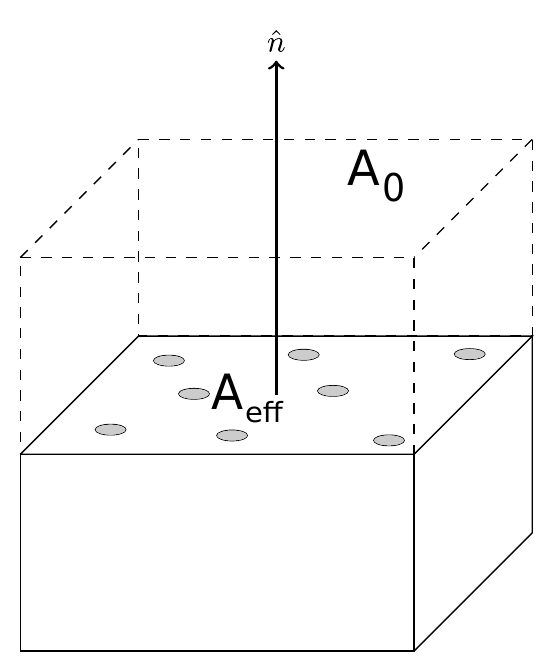}
 \caption{The definition of the damage variable in the model of \citet{Bonora}, where damage is defined as the fractional area of micro-voids intersecting a plane in the representative volume element.}
 \label{fig:DamageVolumeElement}
\end{figure}

\begin{equation}
 D:=1-\frac{A_{\text{eff}}}{A_0} \ .
\end{equation}

Clearly in general $D$ should depend on the normal vector of the intersecting plane, $\hat{n}$, producing a tensorial quantity. However \citet{Bonora} makes the approximation that the distribution of micro-voids can be approximated as isotropic, allowing $D$ to be treated as a scalar field. This approximation greatly simplifies the model, and is valid for highly homogeneous materials. For highly anisotropic materials, such as laminates, this approximation is not appropriate. Instead for such materials a fully tensorial damage model such as \citet{BartonAnisotropicDamage} should be employed. However, this is beyond the scope of this work. Instead, this work seeks to demonstrate a proof-of-concept with a scalar damage model, in the knowledge that more complex damage models could be implemented if desired. In this way, the intention is to provide a general framework that demonstrates the capability of the new numerical method, regardless of what specific constitutive material models are employed.

Therefore, in order to track damage, a scalar field, $D_{(l)}$, for each damageable material is added to the system of equations in the history variable vector:
\begin{align}
\frac{\partial \rho_{(l)}\phi_{(l)} D_{(l)}}{\partial t} + \frac{\partial \rho_{(l)} \phi_{(l)} D_{(l)} u_k}{\partial x_k} = \rho_{(l)}\phi_{(l)}\dot{D}_{(l)} \ .
\end{align}

The inclusion of damage into the thermodynamics of the system employs the method of convex potentials, like plasticity \cite{Barton2019,WallisMultiPhysics,WallisFluxEnriched,WallisRigidBody}. A brief outline of the method is given here, with full details given in previous works \cite{Bonora,WallisFluxEnriched}.
The reduction in load-carrying area caused by the presence of damage leads to a higher effective stress in the material:
\begin{align}
 \widetilde{\boldsymbol{\sigma}} = \frac{\boldsymbol{\sigma}}{1-D} \ ,
\end{align}
where $\widetilde{\boldsymbol{\sigma}}$ is the effective stress and $\boldsymbol{\sigma}$ is the nominal stress that would be present in the absence of damage. This in turn leads to the {\it effective stress} principle first proposed by \citet{kachanov:1958}, also known as the strain equivalence principle, which states that the action of the damaged material under the nominal stress is the same as the action of the undamaged material under the effective stress. This also leads to the assumption that the Young's modulus of the material is also degraded by the presence of damage, further resulting in the linear degradation of the shear and bulk moduli: 
\begin{equation}
{G}_{(l)}=(1-D_{(l)})\widetilde{G}_{(l)},\qquad {K}_{(l)}=(1-D_{(l)})\widetilde{K}_{(l)} \ ,
\end{equation}
where $\widetilde{G}_{(l)},\widetilde{K}_{(l)}$ are the undamaged values.

The contributions to the internal energy from the volumetric and shear strain energies for damageable materials are therefore affected by the damage parameter, and are taken to be 
\begin{align}
\mathscr{E}_{(l)}^c\left(\rho_{(l)},D_{(l)}\right) &= (1-D_{(l)})\widetilde{\mathscr{E}}_{(l)}^{c}\left(\rho_{(l)}\right) \label{eq:coldCompressionDegradation} \\
\mathscr{E}_{(l)}^s\left(\rho_{(l)},\dev{\mathbf{H}^e},D_{(l)}\right) &= (1-D_{(l)})\widetilde{\mathscr{E}}_{(l)}^{s}\left(\rho_{(l)},\dev{\mathbf{H}^e}\right) \ , \label{eq:shearDegradation}
\end{align}
where $\widetilde{\mathscr{E}}_{(l)}^{c}$ and $\widetilde{\mathscr{E}}_{(l)}^{s}$ are the functions corresponding to the undamaged state, and can be any of the forms outlined in previously \cite{Barton2019,WallisMultiPhysics,WallisFluxEnriched,WallisRigidBody}.

In a similar way to plasticity, the Clausius-Dunhem inequality gives the form of the evolution equation for damage in terms of its conjugate thermodynamic force, and the method of convex potentials is then used to derive the evolution of the damage parameter:
\begin{align}
 \dot{D}_{(l)} = - \lambda_{(l)}\pdv{F^D_{(l)}}{Y_{(l)}} \ ,
\end{align}
where $\lambda_{(l)}$ is the Lagrange multiplier, $Y_{(l)} = {\partial\mathscr{E}_{(l)}}/{\partial D_{(l)}}$ is the elastic energy release rate due to damage (the thermodynamic force conjugate to damage), and $F^D_{(l)}$ is the damage dissipation potential, which is a constitutive model that must be defined for a given material, similar to the plastic yield surface. The associative flow rule for damaged materials is:
\begin{equation}
\lambda_{(l)} = \dot{\varepsilon}_{p,(l)}(1-D_{(l)}) \ ,
\end{equation}
where $\varepsilon_{p,(l)}$ is the effective equivalent plastic strain.
The damage dissipation potential is taken from \citet{Bonora}, who propose the following potential to account for micro-mechanical processes:
\begin{align}
 F^D = \left[ \frac{1}{2}\left(-\frac{Y}{S}\right)\frac{S}{1-D_{(l)}}\right] \frac{(D_{\text{crit},(l)}-D_{(l)})^{\frac{\alpha-1}{\alpha}}}{\varepsilon_{p,(l)}^{\frac{2+n}{n}}} \ .
\end{align}
Here $S, \alpha$ and $D_{\text{crit}}$ are material dependent constants, and $n$ is the strain hardening exponent.

From here, the damage rate $\dot{D}$ can be derived. Using the form of the internal energies, the elastic energy release rate is given as
\begin{equation}
-Y_{(l)}  = \rho_{(l)}\widetilde{\mathscr{E}}_{(l)}^{c} + \rho_{(l)}\widetilde{\mathscr{E}}_{(l)}^{s} = \rho_{(l)}\widetilde{\mathscr{E}}_{(l)}^{c} + \widetilde{G}_{(l)}\mathcal{J}^2 \ .
\end{equation}

According to the effective stress principle, the nominal equivalent stress can be expressed as 
\begin{equation}
\sigma_{eq} = \left(1-D_{(l)}\right)\widetilde{\sigma}_{eq} = \left(1-D_{(l)}\right) \sqrt{6} \widetilde{G}_{(l)}\mathcal{J} \ ,
\end{equation}
where $\widetilde{\sigma}_{eq}$ is the effective equivalent stress.

Substituting this into the elastic energy release rate gives:
\begin{equation}
-Y_{(l)}  =  \frac{\sigma_{eq}^2}{\left(1-D_{(l)}\right)^2}\frac{1}{2\widetilde{E}_{(l)}}\left( \frac{2\widetilde{E}_{(l)}\left(1-D_{(l)}\right)^2 \rho_{(l)}\widetilde{\mathscr{E}}_{(l)}^{c}}{\sigma_{eq}^2} + \frac{\widetilde{E}_{(l)}}{3\widetilde{G}_{(l)}}\right) \ ,
\end{equation}
where the Youngs modulus $\widetilde{E}_{(l)}$ has been introduced without loss of generality.

\citet{Bonora} then derives the form for the damage rate:
\begin{align}
 \dot{D}_{(l)} = \alpha \frac{(D_{\text{crit},(l)}-D_{0,(l)})^{\frac{1}{\alpha}}}{\ln(\frac{\varepsilon_{\text{crit}}}{\varepsilon_{\text{thresh}}})}  R_t\left(\frac{p}{\sigma_{\text{eq}}}\right)(D_{\text{crit},(l)}-D_{(l)})^{\frac{\alpha-1}{\alpha}}\left(\frac{\dot{\varepsilon}_p}{\varepsilon_p}\right) \ , \label{eq:DamageSourceTerm}
\end{align}
where $D_{\text{crit}}, D_{0}, \alpha, \varepsilon_{\text{crit}}, \varepsilon_{\text{thresh}}$ are material parameters and $R_t$ is the stress triaxiality function:
\begin{equation}
R_t\left(\frac{p}{\sigma_{\text{eq}}}\right) = \frac{2\widetilde{E}_{(l)}(1-D_{(l)})^2 \rho_{(l)}\widetilde{\mathscr{E}}_{(l)}^{c}}{\sigma^2_{\text{eq}}} + \frac{\widetilde{E}_{(l)}}{3\widetilde{G}_{(l)}} \ .
\end{equation}
It is not immediately apparent that the form of the elastic energy release rate $Y$ matches the well-recognised expressions from the original works of \citet{Bonora,lemaitre:2005} and others, which assume the St. Venant-Kirchhoff model for the strain energy, so here proof is provided. Firstly, the undamaged volumetric strain energy is assumed to be the logarithmic equation of state from \citet{poirier:2000}, which closely resembles the St. Venant-Kirchhoff model:
\begin{equation}
\widetilde{\mathscr{E}}_{(l)}^{c} = \frac{\widetilde{K}_{(l)}}{2\rho_{0,(l)}} \left[\ln \left(\frac{\rho_{(l)}}{\rho_{0,(l)}}\right)\right]^2, \qquad \widetilde{p}_{(l)}^{c} = \widetilde{K}_{(l)} \left(\frac{\rho_{(l)}}{\rho_{0,(l)}}\right) \ln \left(\frac{\rho_{(l)}}{\rho_{0,(l)}}\right) \ ,
\end{equation}
so that 
\begin{equation}
\widetilde{\mathscr{E}}^c_{(l)} = \frac{1}{2\widetilde{K}}\frac{\rho_{0,(l)}}{\rho_{(l)}^2} \left(\widetilde{p}_{(l)}^{c}\right)^2 \ .
\end{equation}
Finally, using the relationships between elastic moduli
\begin{equation}
\frac{\widetilde{E}_{(l)}}{\widetilde{K}_{(l)}} = 3\left(1-2\widetilde{\nu}_{(l)}\right),\qquad \frac{\widetilde{E}_{(l)}}{2\widetilde{G}_{(l)}} = 1+ \widetilde{\nu}_{(l)} \ ,
\end{equation}
where $\widetilde{\nu}$ is the Poisson ratio. Substituting all of the above, the elastic damage energy release rate can be written
\begin{align}
-Y_{(l)} &= \frac{\sigma_{eq}^2}{\left(1-D_{(l)}\right)^2}\frac{1}{2\widetilde{E}_{(l)}} R_t\left(\frac{p^c_{(l)}}{\sigma_{eq}}\right) \\
R_t\left(\frac{p}{\sigma_e}\right) &= \left( 3\left(1-2\widetilde{\nu}_{(l)}\right)\frac{\rho^{(l)}_0}{\rho_{(l)}}\left(\frac{p}{\sigma_{eq}}\right)^2 + \frac{2}{3}\left(1+\widetilde{\nu}_{(l)}\right)\right) \ .
\end{align}
This closely resembles the form outlined by \citet{Bonora}, where instead
\begin{align}
-Y_{(l)} &= \frac{\sigma_{eq}^2}{\left(1-D_{(l)}\right)^2}\frac{1}{2\widetilde{E}_{(l)}} R_t\left(\frac{p}{\sigma_{eq}}\right) \\
R_t\left(\frac{p}{\sigma_e}\right) &= \left( 3\left(1-2\widetilde{\nu}_{(l)}\right)\left(\frac{p}{\sigma_{eq}}\right)^2 + \frac{2}{3}\left(1+\widetilde{\nu}_{(l)}\right)\right) \ .
\end{align}
In this work, the first form of the stress triaxiality is employed, but using the hydrostatic pressure following \citet{Bonora}, rather than the cold compression pressure.

This result proves that this formulation of the damage rate conforms to the strain equivalence hypothesis \cite{lemaitre:2005}, which states that the constitutive model for strain in the damaged state is equivalent to that for undamaged material except the stress is replaced with the effective stress, and the shear and bulk moduli are linearly degraded.

\citet{PirondiAndBorona} extended this model to deal with cyclic loading, proposing the following addition:
\begin{align}
 \varepsilon_p^- &= \left\lbrace\mqty{ \varepsilon_p && \text{if: } \ p < 0 \\ 0 && \text{else} \\}\right. \\
 \dot{D} &= \left\lbrace\mqty{ \dot{D} && \text{if: } \ p < 0 \ \text{and} \ \varepsilon_p^- > \varepsilon_{p,\text{thresh}} \\ 0 && \text{else} \\}\right.
\end{align}
which conveys the underlying assumption that damage can only accrue in states of tension, not under compression. Finally, threshold plastic strain $\varepsilon_{p,\text{thresh}}$ is taken to depend on the stress triaxiality using the model from \citet{StressTriaxialityDependence}:
\begin{align}
 \varepsilon_{p,\text{thresh}} = \frac{\varepsilon_{\text{thresh}}^{1/R_t}}{R_t^{1/n}}\left( \frac{\varepsilon_{\text{crit}}^{2n}-\varepsilon_{\text{thresh}}^{2n}}{\varepsilon_{\text{crit}}^{2n/R_t}-\varepsilon_{\text{thresh}}^{2n/R_t}}     \right)^{\frac{1}{2n}} \ ,
\end{align}
where $n$ is the strain hardening exponent.

\subsubsection{Closure Models}
\label{sec:ClosureModels}

When considering elastoplastic solids, both the volumetric and shear strain energy closure models must be provided. For the volumetric strain energy, this work follows previous studies by using the form outlined by \citet{RomenskiiEOS}:
\begin{align} \label{eq:Romenskii}
\mathscr{E}_{(l)}^c = \frac{K_{0,(l)}}{2\rho_{(l)}\alpha_{(l)}^2}\left(\left(\frac{\rho_{(l)}}{\rho_{0,(l)}}\right)^{\alpha_{(l)}}-1\right)^2 \ , 
\end{align}
where $\alpha,K_{0}$ and $\rho_0$ are material parameters.
The shear energy always takes the form in equation \ref{eq:shearEnergy}, but the form of the shear modulus $G$ is a closure model. This work considers two options for $G$. The first is the form outlined by \citet{RomenskiiEOS}:
\begin{align}
G_{(l)}(\rho_{(l)}) &= G_{0,(l)}\left(\frac{\rho_{(l)}}{\rho_{0,(l)}}\right)^{\beta_{(l)}+1} \ ,
\end{align}
where $G_{0}$ and $\beta$ are material parameters.
The second is the form outlined by \citet{SteinbergEOS}:
\begin{align}
G_{(l)}(\rho_{(l)}) &= \left(G_{0,(l)} + G_{p,(l)}p_c(\rho)\left(\frac{\rho}{\rho_0}\right)^{-1/3} \right) \ ,
\end{align}
where $p_c(\rho) = \rho^2\pdv{\mathscr{E}_{(l)}^c}{\rho}$ is the cold-compression pressure and $G_p$ is a material parameter.
When the \citet{RomenskiiEOS} shear energy is used, the Debye temperature and Gr\"{u}neisen function are given by:
\begin{align}
\theta_{(l)}^{\text{D}}(\rho_{(l)}) &= \left(\frac{\rho_{(l)}}{\rho_{0,(l)}}\right)^{\Gamma_{0,(l)}} \\
\Gamma_{(l)}(\rho_{(l)}) &= \Gamma_{0,(l)} \ , 
\end{align}
where $\Gamma_0$ is a material parameter.
When the \citet{SteinbergEOS} shear energy is used, the Debye temperature and Gr\"{u}neisen function are given by form from \citet{BurakovskyPreston}:
\begin{align}
\theta_{(l)}^{\text{D}}(\rho_{(l)}) &= \left(\frac{\rho_{(l)}}{\rho_{0,(l)}}\right)^{\Gamma_{\infty,(l)}}\exp\left[3\Gamma_{1,(l)}\left(1-\left(\frac{\rho_{(l)}}{\rho_{0,(l)}}\right)^{-1/3}\right) + \frac{\Gamma_{2,(l)}}{\gamma_{(l)}}\left(1-\left(\frac{\rho_{(l)}}{\rho_{0,(l)}}\right)^{-\gamma_{(l)}}\right) \right] \\
\Gamma_{(l)}(\rho_{(l)}) &= \Gamma_{\infty,(l)}+\Gamma_{1,(l)}\left(\frac{\rho_{(l)}}{\rho_{0,(l)}}\right)^{-1/3}+\Gamma_{2,(l)}\left(\frac{\rho_{(l)}}{\rho_{0,(l)}}\right)^{-\gamma_{(l)}} \ , 
\end{align}
where $\Gamma_{\infty},\Gamma_1,\Gamma_2 $ and $\gamma$ are material parameters.

Finally, when reactive fluids are considered, the Jones-Wilkins-Lee (JWL) equation of state is employed for both the reactants and products of the explosive. For this equation of state $\Gamma(\rho)=\Gamma_0$ and 
\begin{align}
  \mathscr{E}_{\text{\scriptsize{ref}},(l)}(\rho_{(l)}) &= \frac{{\cal A}_{(l)}}{{\cal R}_{1,(l)}\rho_{0,(l)}}e^{-{\cal R}_{1,(l)}\frac{\rho_{0,(l)}}{\rho_{(l)}}}+\frac{{\cal B}_{(l)}}{{\cal R}_{2,(l)}\rho_{0,(l)}}e^{-{\cal R}_{2,(l)}\frac{\rho_{0,(l)}}{\rho_{(l)}}} \\ 
  p_{\text{\scriptsize{ref}},(l)}(\rho_{(l)}) &= \rho^{2}_{(l)} \frac{\partial\mathscr{E}_{\text{\scriptsize{ref}},(l)}}{\partial \rho_{(l)}} = {\cal A}_{(l)}e^{-{\cal R}_{1,(l)}\frac{\rho_{0,(l)}}{\rho_{(l)}}}+{\cal B}_{(l)}e^{-{\cal R}_{2,(l)}\frac{\rho_{0,(l)}}{\rho_{(l)}}}
\end{align}

Having outlined the single-material closure models, these must then be combined in mixture regions in a thermodynamically consistent way. Mixture rules designed for this task are given in previous studies \cite{Allaire,MiNi16,Barton2019,WallisMultiPhysics,WallisFluxEnriched}.

\subsubsection{Lagrangian Perturbation Field}
\label{sec:PerturbationField}
\begin{figure}
\centering
\includegraphics[width=0.9\textwidth]{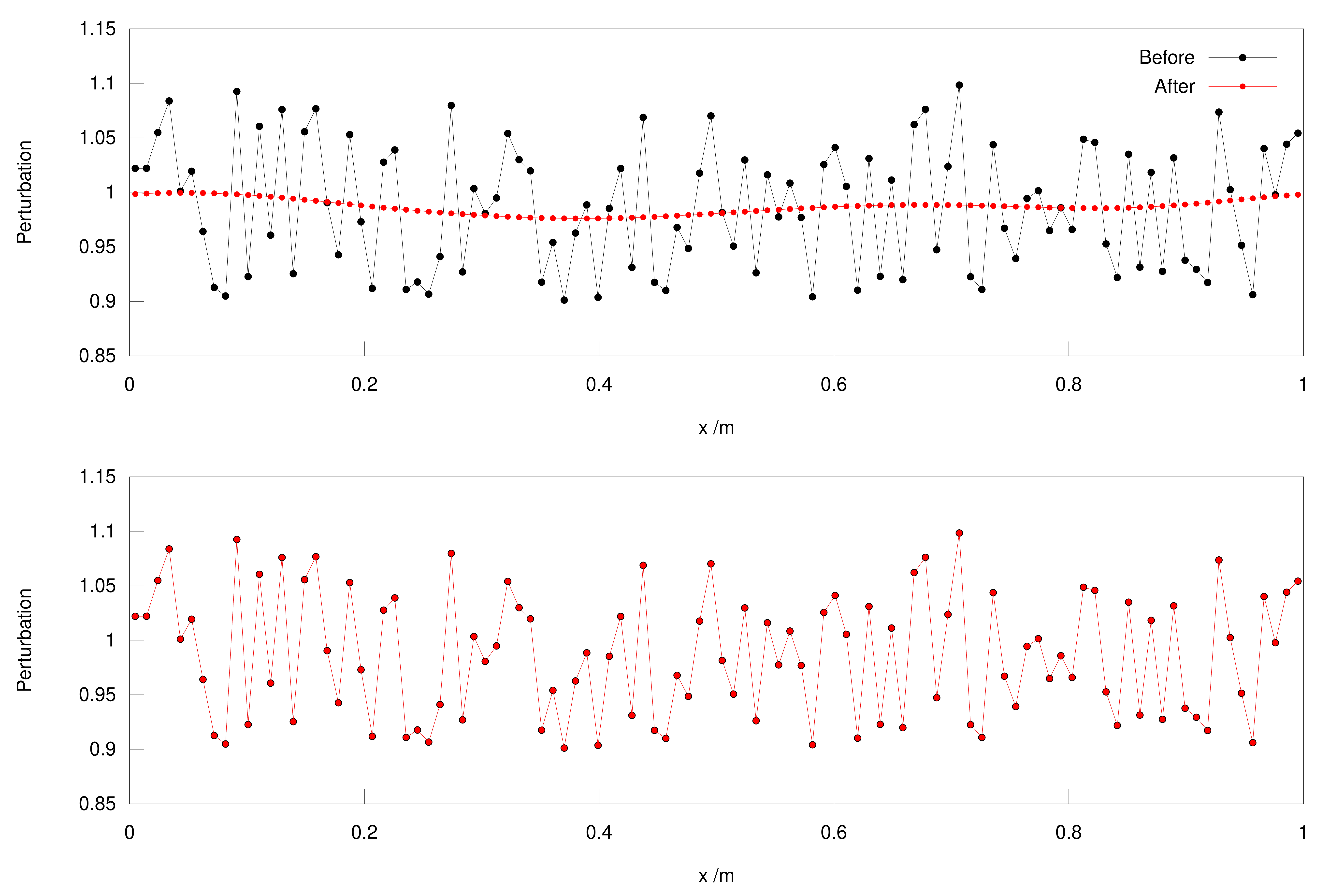}
\caption{The advection of the perturbation field after evolving one domain-length. (\textit{Top}) Using Eulerian methods. (\textit{Bottom}) Using the semi-Lagrangian update of \citet{VitaliTransportDiffusion}. Clearly Eulerian methods are not suited to this task, as they diffuse the profile too severely. The semi-Lagrangian update avoids this issue.}
\label{fig:PerturbationFieldAdvection}
\end{figure}

This work models random material inhomogeneities by perturbing the constitutive models. This method has been used by a number of authors to account for micro-scale material detail \cite{VitaliTransportDiffusion, RabczukPerturbation, BeckerRingFracture, BartonLevelSetDamage}. This method evolves a scalar field, $\varphi$, that acts as another material history variable, perturbing the material yield stress and resulting in the localised failure that is desired. Crucially, however, were this field to be advected using a purely Eulerian scheme, numerical diffusion would quickly smear out the high-frequency modes of any random distribution, rendering the field practically useless, as shown in Figure \ref{fig:PerturbationFieldAdvection}. 

One solution to this problem would be to employ an extremely high order numerical method, but a more straightforward way is the semi-Lagrangian method proposed by \citet{VitaliTransportDiffusion}, subsequently also employed by \citet{BartonLevelSetDamage} for level-set based fracture. In this scheme, $\varphi$ is tracked by evolving the material coordinates, $\mathbf{X}$, in the Eulerian frame, $\mathbf{x}$, and mapping the field $\varphi$ from the material reference frame to the Eulerian coordinates as required.

Therefore, equations of motion for the material coordinates $X_i$ in the Eulerian frame must be added to the system of equations:
\begin{align}
 \dot{X_i} &= 0 \\
 \pdv{X_i}{t} + \pdv{X_iu_k}{x_k} &= 0 \ .
\end{align}

The initial perturbation field is defined in the initial material coordinate frame $\varphi_0(\mathbf{X}_0)$, where both frames are identical: $ \mathbf{X}_0 = \mathbf{X}(x,t=0) = \mathbf{x}_0$. Subsequently, the Eulerian value of $\varphi(x,t)$ is obtained from the new Lagrangian coordinates by interpolating values from the initial perturbation field at $\mathbf{X}(x,t)$:
\begin{align}
 \varphi(x,t) = \varphi_0(\mathbf{X}(x,t)) \ .
\end{align}

The material coordinates are relatively smoothly varying functions of space that can updated straightforwardly and diffusion in the perturbation field is completely avoided, without resorting to very high order methods. 

Once the Eulerian cell-centre value of $\varphi$ has been obtained, this value then multiplies the yield stress in the plastic source term:
\begin{align}
 \sigma'_Y\left(\varepsilon_{p,(l)},\varphi\right) = \varphi \sigma_Y\left(\varepsilon_{p,(l)}\right)\ .
\end{align}

This is then found to be sufficient to initiate the failure localisation required.

\section{Numerical Approach}

The numerical approach for the update of the thermodynamic variables and the treatment of plasticity has been covered in detail in previous works \citet{Barton2019, WallisMultiPhysics, WallisFluxEnriched, WallisRigidBody}, so shall not be repeated here. It only remains to describe the method for fracture and the update of the Lagrangian fields.

\subsection{Fracture}

A fourth-order Runge-Kutta integration (RK4) is used to update the damage source term. Fracture is then initiated using exactly the same void-opening flux modifier as presented in \citet{WallisFluxEnriched}. It is now applied when either cell between which the flux is to be calculated, $\mathbf{U}_L$, $\mathbf{U}_R$, is critically damaged, and the cells are under tension in the hyperbolic sweep direction currently being considered. For example in the $x$-direction:
\begin{align}
\mathbf{\hat{x}}^{\text{T}} \cdot \left(\boldsymbol\sigma_R +\boldsymbol\sigma_L\right) \cdot \mathbf{\hat{x}} > 0 \ .
\end{align}

Fracture is handled naturally by the same routines that mediate interface separation. Critically damaged cells, defined as any cell where:
\begin{equation}
D_{(l)} \ge D_{\text{crit},(l)} \ ,
\end{equation}
are simply treated as any other material interface where the void-opening flux-modifier can be applied. No additional algorithm needs to be applied to initiate fracture. This approach also removes the need for the highly non-conservative approach of deleting critically damaged cells to form fracture, as is commonplace in finite element and level set based codes \cite{ParticleLevelSetDamage, UdaykumarDamage, BartonLevelSetDamage}. This also avoids the issue of potentially having to redistribute the conserved quantities such a mass and momentum to the neighbouring cells around the deleted cells. 

However, this method does come with its own associated challenges; the diffusive nature of shock-capturing Eulerian codes means that damaged cells may undergo artificial healing. As an initially critically damaged material is advected, diffusion can `heal' the damage by causing it to fall under the critical damage threshold. This causes a number of issues:

\begin{itemize}
 \item Regions that should be able to come apart and form a crack are prevented from separating.
 \item The transition to critically damaged and back is discontinuous; quantities such as the yield strength go suddenly to zero.
 \item When a history-dependent plasticity model such as strain-hardening is employed, the plastic strain remains unchanged through healing. This means a material can be damaged, heal, and then continue to accrue plastic strain, leading to unphysical increased hardening. 
\end{itemize}

Additionally, as the measure of whether two cells are under tension is taken along the hyperbolic sweep direction, this introduces some grid-dependence into the void-generation associated with damage. This is because damage does not necessarily relate to an identifiable interface, so an interface normal cannot be calculated for the tension criterion. However, this could be remedied by employing a tensorial damage model, such that the direction of the damage interface could be ascertained. Moreover, the material inhomogeneities produced by the damage perturbation field dominate this effect in the examples considered in this work.

\subsection{Modelling Localised Failure with Lagrangian Perturbation Fields}

As has been outlined above, this work includes a randomly-varying scalar field $\varphi$ to model the effect of real material inhomogeneities and produce realistic fragment distributions. The method consists of two elements: the advection of the Lagrangian coordinates and the interpolation of the scalar field value. The underlying method for both parts is largely unchanged from that presented in \citet{VitaliTransportDiffusion} and \citet{BartonLevelSetDamage}, and so shall only be summarised here.

\subsubsection{Lagrangian Coordinate Evolution}

During each time step, the Lagrangian coordinates, $\Lagrangian(x,t)$, are updated concurrently with the hyperbolic update, using a third order Runge-Kutta (RK3) time integration. Each time step uses an upwind method with third order WENO \cite{WENO} reconstruction.

The equations of motion are already in conservation law form, so the same discretised conservative approximation as the hyperbolic update can be used:
\begin{equation} 
  \frac{\dd}{\dd t}{\mathbf{U}}_{ijk}^n+\mathcal{D}_{ijk}^n\left({\mathbf{U}}\right) = 0,
\end{equation} 
where ${\mathbf{U}}_{ijk}^n = \Lagrangian_{ijk}^n$ represents the vector of Lagrangian coordinates stored at cell centres,  and 
\begin{eqnarray}
 \mathcal{D}_{ijk}^n :=&& \frac{1}{\Delta x_{ijk}}\left({\mathbf{F}}^n_{i+1/2,jk}-{\mathbf{F}}^n_{i-1/2,jk}\right)\nonumber\\
&+&\frac{1}{\Delta y_{ijk}}\left({\mathbf{G}}^n_{i,j+1/2,k}-{\mathbf{G}}^n_{i,j-1/2,k}\right)\nonumber\\
&+&\frac{1}{\Delta z_{ijk}}\left({\mathbf{H}}^n_{ij,k+1/2}-{\mathbf{H}}^n_{ij,k-1/2}\right)
\end{eqnarray}
where ${\mathbf{F}}, {\mathbf{G}}, {\mathbf{H}}$ are the cell-wall numerical flux functions. \citet{BartonLevelSetDamage} employs a simple first order upwind scheme for the numerical flux functions. However, this work uses a third order upwind WENO update, as this was found to provide greater accuracy in some cases, especially those involving thin geometries. 

One additional modification is made to the method in order to be compatible with the scheme at hand. The method presented in \citet{VitaliTransportDiffusion} assumes that a velocity field is present across the entire domain, with which the Lagrangian coordinates can be updated. For the scheme at hand, however, the localised nature of the interface seeding routines around void boundaries means that the material velocity is only seeded into a small region of void around a material. This means that Lagrangian coordinates far away from the material are not suitably updated. An interface seeding method is employed to alleviate this issue, completely analogously to those previously presented \cite{WallisFluxEnriched, WallisRigidBody}. This seeding fills the cells around the material with extrapolated Lagrangian coordinates, enabling the Lagrangian coordinates around the material to be suitably updated with the standard method. As has been mentioned previously \cite{WallisFluxEnriched}, this work attempts to emulate many of the techniques of the ghost-fluid method \cite{FedkiwGFM} in a diffuse interface context. By way of example, an analogous extrapolation technique was performed by \citet{BartonLevelSetDamage}, where the Lagrangian coordinates were extrapolated out from the interface using a level set method. \\

%\subsubsection{Lagrangian Seeding Routine} %%%%%%%%%%%%%%%%%%%%%%%%%%%%%%%%%%%%%%%%
%%%%%%%%%%%%%%%%%%%%%%%%%%%%%%%%%%%%%%%%%%%%%%%%%%%%%%%%%%%%%%%%%%%%%%%%%%%%%%%
%%%%%%%%%%%%%%%%%%%%%%%%%%%%%%%%%%%%%%%%%%%%%%%%%%%%%%%%%%%%%%%%%%%%%%%%%%%%%%%

\framebox[\textwidth]
{
\begin{minipage}{0.9\textwidth}
{
The Lagrangian seeding routine:
\begin{enumerate}
\item The seeding is performed in void ($\nu > \nu_{\text{Thresh}}$) cells. The normal vector $\normal_{\nu}$ is calculated with the void volume fraction. The cell centre position of the cell to be seeded is denoted $\mathbf{c}$.
\item A probe is sent out along the normal direction a distance $1.5 \ \dd x$ and the Lagrangian coordinates $\Lagrangian(x,t)$ are interpolated at that point, $\mathbf{p}$, giving $\Lagrangian_{\text{interp}}$.
 \item The new Lagrangian coordinates are then given by:
 \begin{equation*}
  \Lagrangian_{\text{new}} = \Lagrangian_{\text{interp}} - ( \mathbf{p} - \mathbf{c}) \ ,
 \end{equation*}
 where $\mathbf{p}$ and $\mathbf{c}$ are the probe position and the cell centre position respectively.
\end{enumerate}
}\end{minipage}}\\[0.5cm]

\subsubsection{Interpolating the Initial Random Field}

The random scalar field is initialised with any desired distribution, generally with a mean value of 1. This field then multiplies the yield stress of the material, producing the effect of locally raising or lowering the strength of the material. This has the effect of mimicking defects or inhomogeneities in real materials, enabling better analysis of fragmentation distributions. 

Two different random distributions are considered in this work.
\begin{itemize}
 \item The uniform distribution:
 \begin{equation}
 \varphi \sim \mathcal{U}(1-a,1+a)
 \end{equation}
 where any value in $[1-a:1+a]$ has an equal probability of being chosen. Here $a<1$ is a parameter representing the size of the variation.
 \item The rectified Gaussian/normal distribution:
 \begin{equation}
 \varphi \sim \mathcal{N}^R(\mu=1,\sigma^2)
 \end{equation}
 where $\mu$ is the mean, taken to be 1, and $\sigma$ is the standard deviation. The rectification ensures no field values are negative by setting any would-be negative values to 0.
\end{itemize}

As previously mentioned, the field could also be initialised with an empirically determined distribution for a given application of interest. However, this work attempts to provide a proof-of-concept for the method in the current context, rather than modelling a specific application.

The initial perturbation field can be defined on any desired mesh, not necessarily conforming to the Eulerian material mesh, providing that it completely covers any damageable solid bodies of interest. The Eulerian cell-centre field value is required during the plastic source term update. However, as only the initial field is stored, this must be obtained by interpolating the value from the initial field. The interpolation position is given by the evolved Lagrangian coordinates currently held in the Eulerian cell centre in question.

There are several options for the storage of the initial scalar field for parallel distributed memory programs. Either every processor can be given a full copy of the initial mesh, which removes the need for parallel communication but introduces a significant memory burden for large applications, or the initial field can be stored in a distributed fashion with the necessary communication overhead. Alternatively the smaller `material mesh' method employed by \citet{BartonLevelSetDamage} can be used, where, when the material volume is small relative to the domain size, the initial Lagrangian mesh is defined only in a compact volume around any material in question. This avoids excessive memory burden, and a copy can be passed to each processor. This last option is chosen for this work.

\section{Validation and Verification}

In this section, the new damage and fracture model is validated in one, two and three dimensions.

\subsection{One dimensional spallation fracture}

The new damage model is first validated in a quasi-one-dimensional test based on the experimental spallation test from \citet{1DSpallationPaper}. In this test, a block of aluminium alloy 5083 H32 collides with a block of copper alloy CuBe TF00 (C17200), causing the copper to undergo spallation fracture. Spallation damage is an important test for any fracture model, as it requires void-generation completely inside one material, which cannot physically be modelled by the artificial injection of a low density gas, as done by some other methods. This is therefore a good candidate for validation, and the results can be compared with the experimental back-surface velocity profile measured by \citet{1DSpallationPaper}.

\begin{figure}
\centering
\includegraphics[width = 0.6\textwidth]{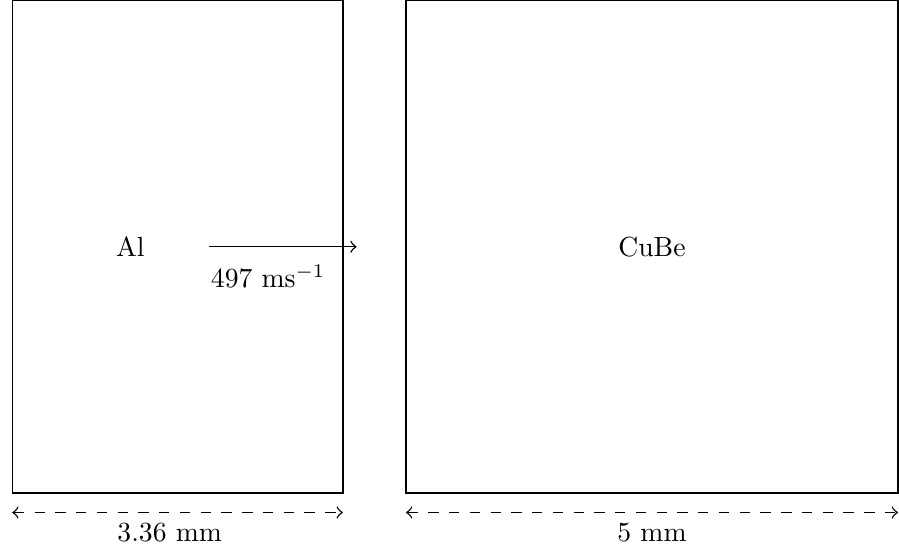}
\caption{The 1D spallation test initial conditions.}
\label{fig:1DSpallationIntialConditions}
\end{figure}

The initial conditions for the test are shown in Figure \ref{fig:1DSpallationIntialConditions}. The test is run for 5 $\mu$s on a domain spanning $x = [0:11]$ mm, $y=[0:11]$ mm, using a CFL of 0.4 with a varying number of cells, but with the same number of cells in each direction.

Exact material parameters for the materials in question were not available for this test, but can be approximated from the data present in \citet{1DSpallationPaper} and \citet{Bonora}. Both materials are governed by the \citet{RomenskiiEOS} equation of state, with parameters given in Table \ref{tab:DamageRomenskiiParameters}. Both materials obey Johnson Cook plasticity, with parameters laid out in Table \ref{tab:DamagePlasticParameters}. The copper is damageable, with parameters laid out in Table \ref{tab:DamageParameters}.

\begin{figure}
\centering
\includegraphics[width = \textwidth]{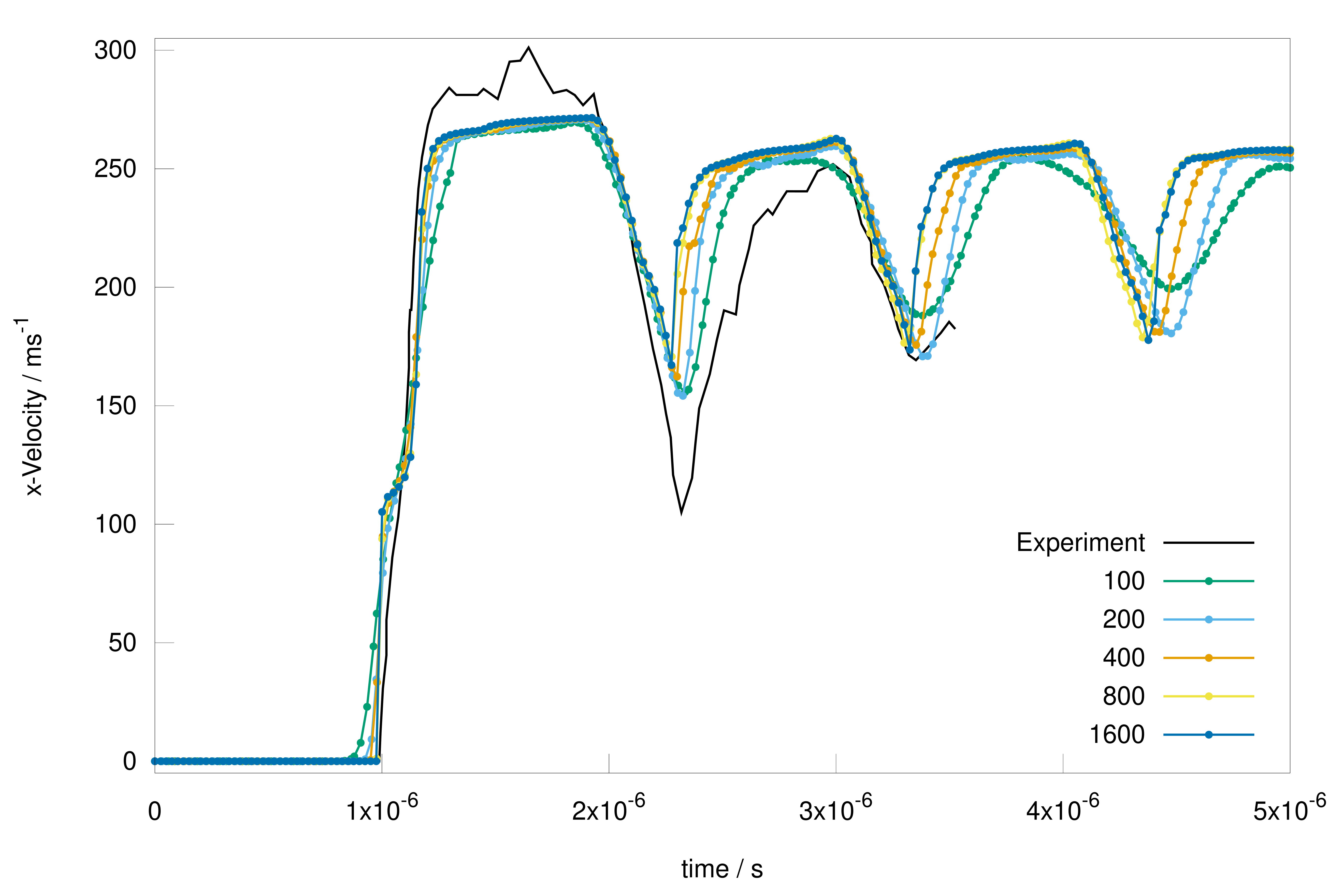}
\caption{The back-surface velocity profile for the spallation fracture test, comparing to experiment. This test demonstrates a combination of damage, fracture and void generation working well in one dimension.}
\label{fig:1DSpallation}
\end{figure}

This test can be perform both with and without the Lagrangian field for comparison. Initially, the test is performed without the Lagrangian field over a number of resolutions and the results are shown in Figure \ref{fig:1DSpallation}. The results agree well with those of \citet{1DSpallationPaper}, with the pull-back strength and oscillation frequency matching closely. However, it should be noted that the strength of the pull-back in the test is very sensitive to the material parameters chosen.

Next, this test measures the effect of varying the Lagrangian field on the resulting fragment distributions, with similar tests being performed by \citet{VitaliTransportDiffusion}. The same initial conditions and material parameters are used as before, with a resolution of 800 $\times$ 800 cells, but the copper is also given a Lagrangian field damage perturbation to model material inhomogeneities. In this case, the field has a uniform distribution in $[1-a:1+a]$, where $a$ is a parameter.

\begin{figure}
\centering
\includegraphics[width = \textwidth]{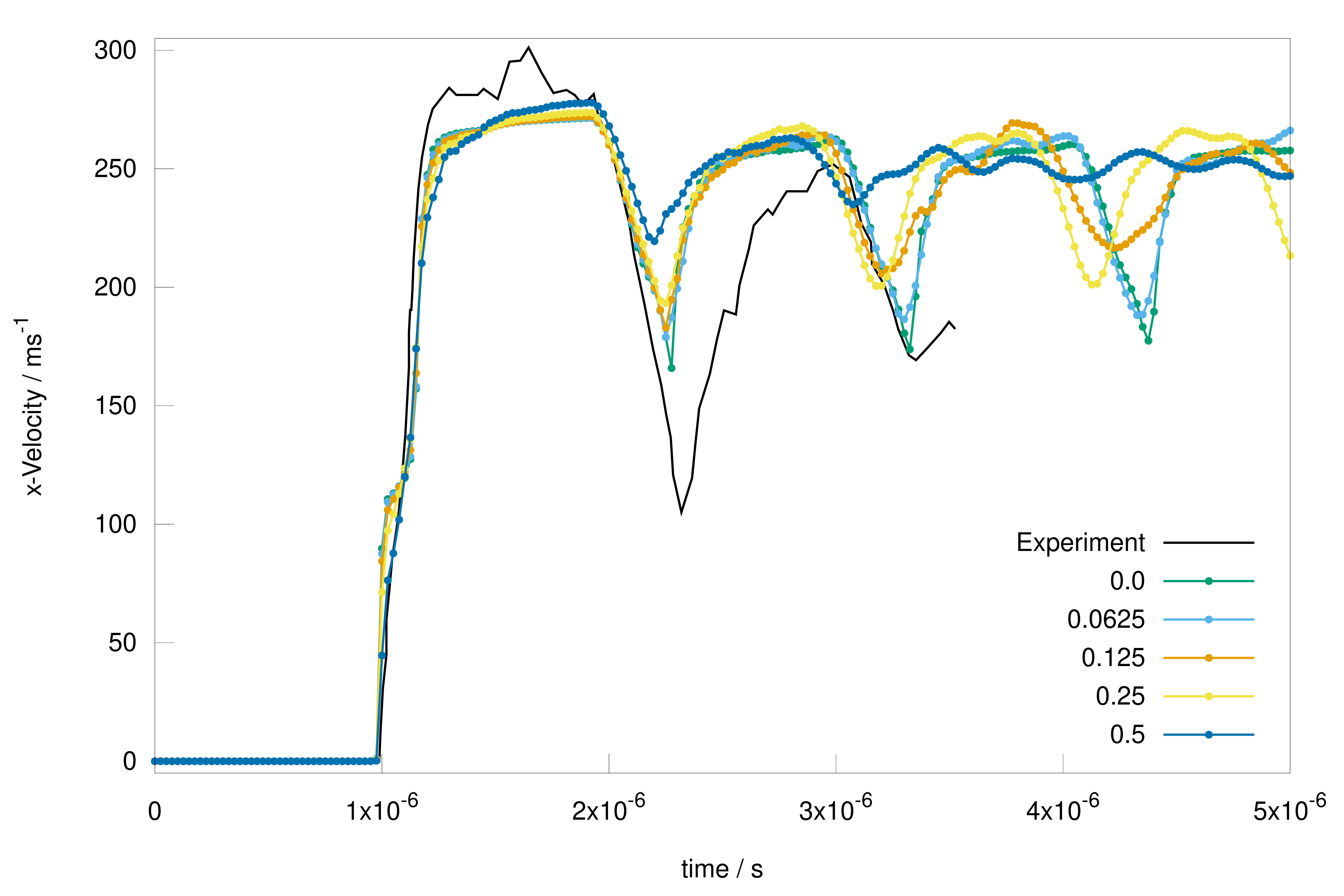}
\caption{The back-surface velocity profile of the spallation fracture test with the Lagrangian field. This plot compares the numerical back surface velocity profile for different amplitudes of perturbation to the experimental profile. The presence of the randomly varying field strongly damps the oscillations.}
\label{fig:SpallationFractureVelocity}
\end{figure}

\begin{figure}
\centering
\includegraphics[width = \textwidth]{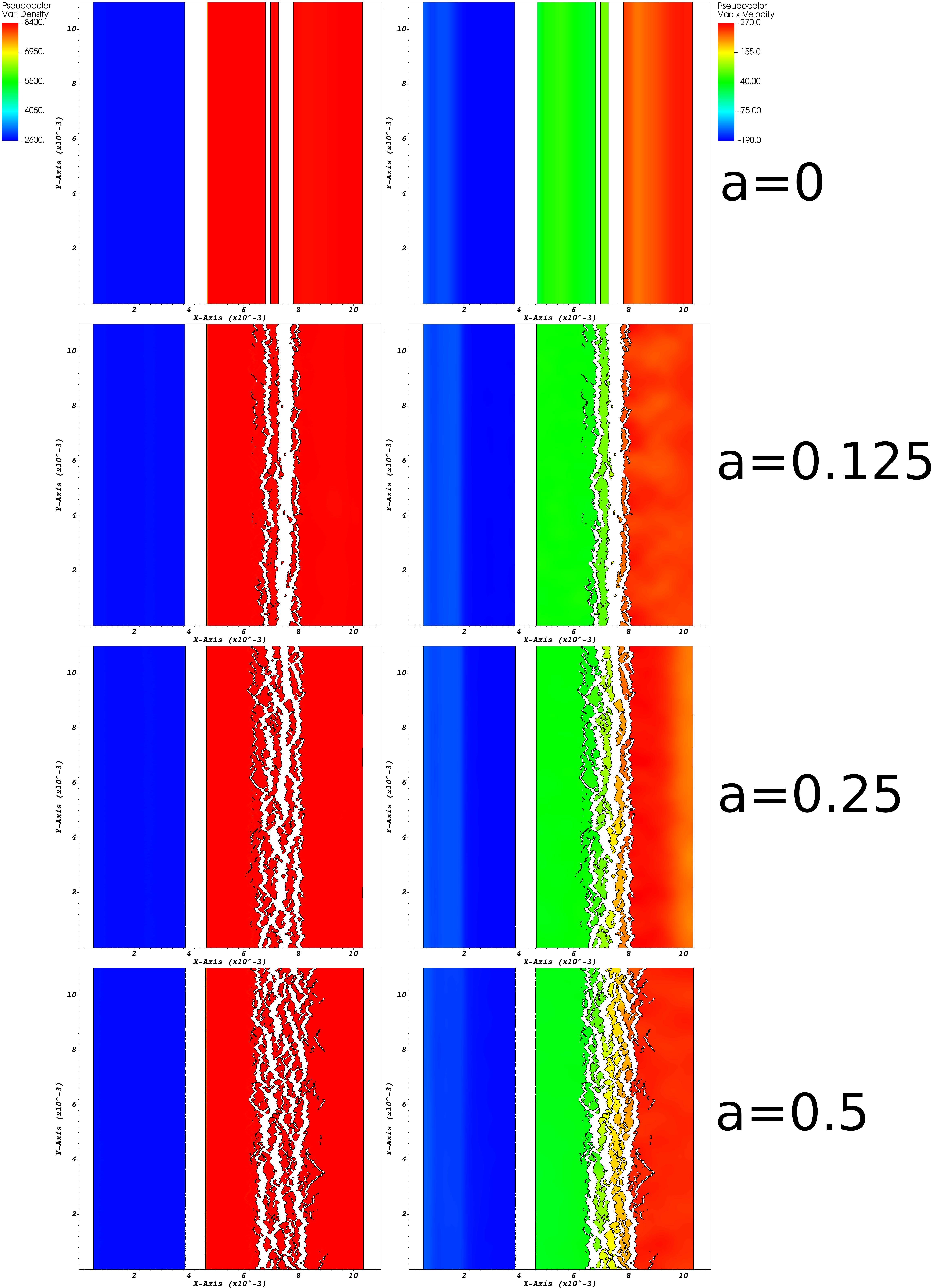}
\caption{The spallation fracture test with the Lagrangian field. The images show (\textit{left}) the density, (\textit{right}) the $x$-velocity at a time of 5 $\mu$s. The Lagrangian field perturbation amplitude $a$ increases from top to bottom.}
\label{fig:SpallationFracture}
\end{figure}

The test is shown in Figure \ref{fig:SpallationFracture}, where the density and $x$-velocity are shown at a time of 5 $\mu$s for a range of different perturbation amplitudes. As the perturbation increases, fragments become larger and more numerous, but the overall location of the spallation remains constant. The experimental comparison for this test is shown in Figure \ref{fig:SpallationFractureVelocity}. Here the effect of the damage perturbation can be seen quantitatively, as the increasing perturbation size strongly damps the oscillations in the velocity.

\subsection{Expanding disk}

This test measures the effect of resolution on the fracture of an expanding tensile disk. This test is similar to that in \citet{OwenExpandingDisk}. The test is two dimensional, consisting of a circle of 304L stainless steel, radius 10 cm, centred at the origin and surrounded by vacuum. The metal is given an initial velocity distribution to initiate fracture. In this case, the velocity distribution is $u(r) = 2000\vec{r}$ ms$^{-1}$. These conditions produce a strenuous test of the model; rather than featuring a single spallation crack as in the previous test, the tensile disk fractures throughout its entire area, producing complex branching cracks. The sheer amount of damage and number of fragments produced in this test makes it a good candidate for examining the behaviour under mesh refinement.

The steel is governed by the \citet{RomenskiiEOS} equation of state with parameters given in Table \ref{tab:DamageRomenskiiParameters}. The steel uses the Johnson-Cook plasticity law and the \citet{Bonora} damage model, with the parameters given in Tables \ref{tab:DamagePlasticParameters} and \ref{tab:DamageParameters}. These parameters are taken from various sources in the literature \cite{Maurel-Pantel_304LSS_Plasticity,Wang_304LSS_Damage,CampbellExplosiveShell}.

The steel is given a Lagrangian field damage perturbation as before. In this case a Gaussian distribution is used, with a mean of 1.0 and a standard deviation of 0.1.

The test is run for 200 $\mu$s using a CFL of 0.4, with a domain spanning $x = [-13:13], y = [-13:13]$ cm. Four different resolutions are tested: $200\times200, 400\times400, 800\times800$ and $1600\times1600$.

\begin{figure}
\centering
\includegraphics[width = \textwidth]{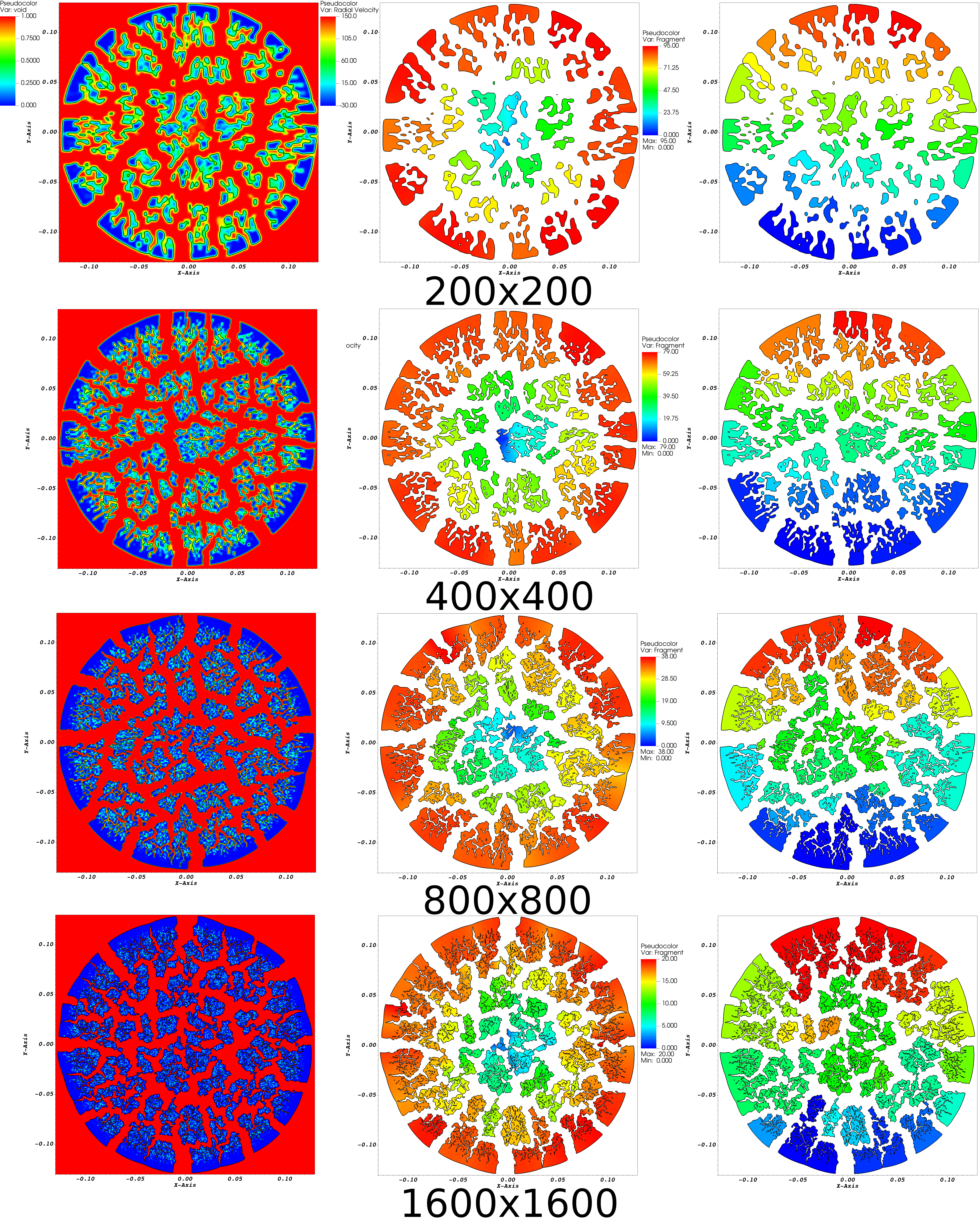}
\caption{The expanding disk experiment. The images show the four different resolutions tested from top to bottom, and the void volume fraction, radial velocity and fragment number from left to right. The images are all taken at 200 $\mu$s.}
\label{fig:Disk}
\end{figure}

\begin{figure}
\centering
\includegraphics[width = \textwidth]{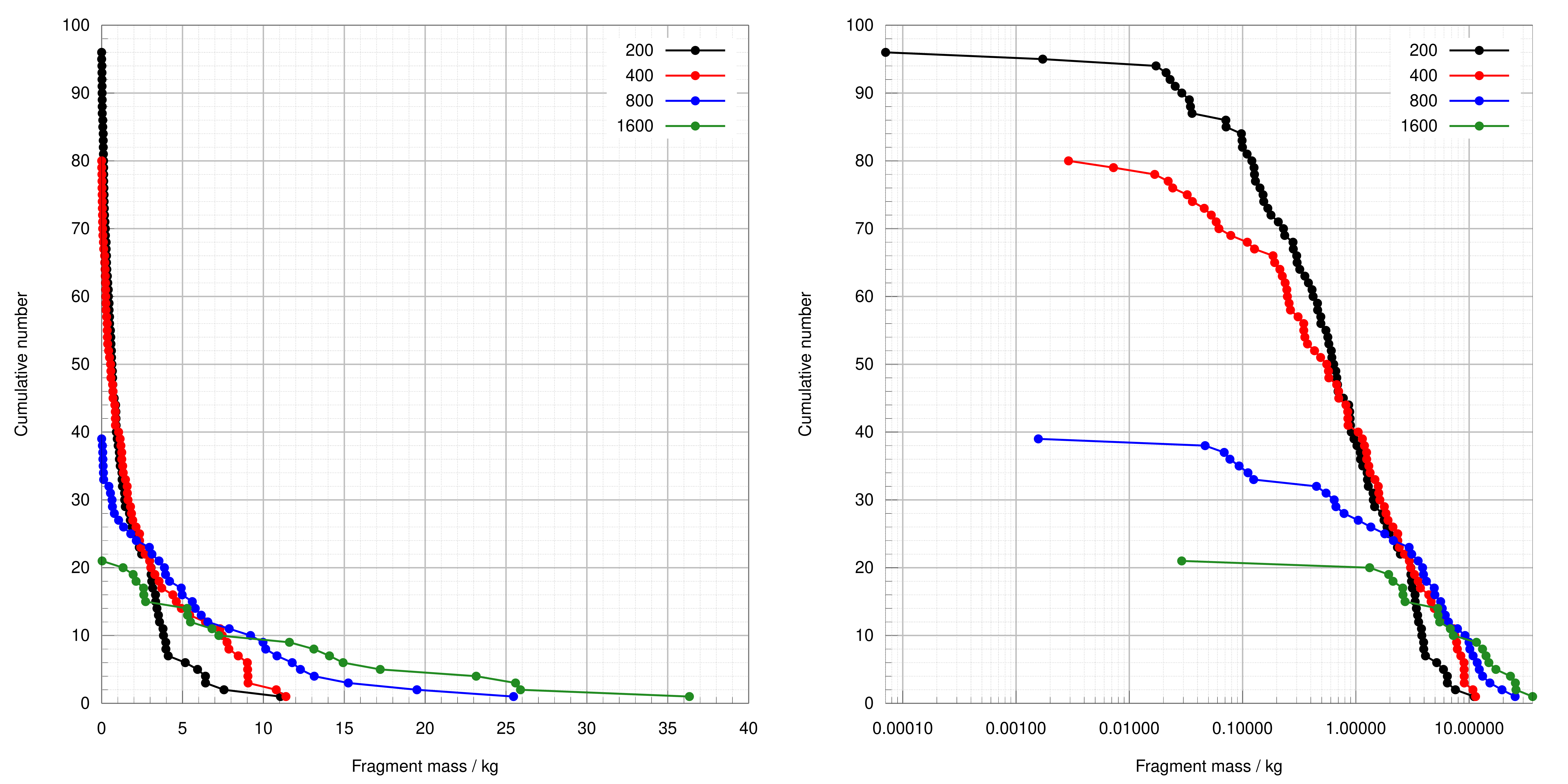}
\caption{The fragment distributions in expanding disk experiment. The images show the fragment mass against the cumulative fragment number on (\textit{left}) linear and (\textit{right}) logarithmic scales at 200 $\mu$s.}
\label{fig:DiskFragments}
\end{figure}

Figure \ref{fig:Disk} shows the comparison of the different resolutions. The image depicts the void volume fraction, radial velocity, and fragments for each resolution tested. In all the images, fragments are demarcated by the 0.5 void volume fraction contour. The test performs well, producing decent fragments even at very low resolution over this long test. The different fragment distributions produced in the test are compared in Figure \ref{fig:DiskFragments}.

\subsection{Expanding ring}

This test considers a ring driven radially outward until fracture. \citet{MottRingFracture} was one of the first to consider this kind of test, developing a theoretical estimate of the fragment mass distribution. \citet{MottRingFracture} produced a statistical estimate for the average fragment length by relating the variation of the strength of the material to the chance a fracture will form, and then considering the propagation of stress release waves in the ring after fragmentation which prevent further damage. Subsequently, there have then been several experimental \cite{NiordsonRingFracture, GourdinRingFracture}, and numerical \cite{BeckerRingFracture, ZhangRingFracture, MeulbroekRingFracture, BartonLevelSetDamage} studies examining this test. The combination of theoretical, experimental and numerical studies makes this test an excellent candidate for validation.

This work examines the test as it is presented in \citet{BartonLevelSetDamage}, in which a ring of U-Nb alloy is driven radially outward until fracture. Previous experiments have tried a variety of ways to propel rings to fracture, such as using gas-guns or explosives, but this work focuses on electromagnetically driven rings. A solenoid produces a magnetic field that pushes a driver ring outwards, which in turn pushes a specimen ring. A U-Nb alloy was chosen for the specimen due to its low self-inductance, such that it would be predominantly affected by the driver ring, and not the field produced by the solenoid. This work mimics the experiment by driving the ring with an imposed velocity profile that matches the experimental work by use of the rigid body method set out in \citet{WallisRigidBody}.

The ring has an outer radius of 17.945 mm, an inner radius of 17.185 mm and a thickness of 0.76 mm, resulting in a square cross-section \cite{BartonLevelSetDamage}. The velocity is chosen to approximate the 6 kV profile given by \citet{BeckerRingFracture}:

\begin{align}
 v_{r} = \left\lbrace\mqty{ v_0 \frac{t}{t_0} && \text{if } t < t_0 \text{ and } r < r_{\text{\scriptsize{max}}} \\ v_0 + (v_1 - v_0)\frac{t-t_0}{t_1-t_0} && \text{if } t_0 < t < t_1 \text{ and } r < r_{\text{\scriptsize{max}}} \\ v_1 && \text{if } t_1 < t \text{ and } r < r_{\text{\scriptsize{max}}} \\ 0 && \text{if } r > r_{\text{\scriptsize{max}}}}\right. \ ,
\end{align}
where $v_0$ = 210 ms$^{-1}$, $v_1$ = 160 ms$^{-1}$, $t_0$ = 17.5 $\mu$s, $t_1$ = 35 $\mu$s, $r_{\text{\scriptsize{max}}}$ = 20 mm. The parameter $r_{\text{\scriptsize{max}}}$ is chosen to include the effect of the arrestor ledge mentioned in the experimental work that halts the driving at 20 mm, letting the ring continue freely. This approximation to the experimental velocity was found to be sufficient for the purposes of this study, but a more accurate velocity profile could be implemented if desired. Following a similar technique in \citet{BartonLevelSetDamage}, the rigid body was fully removed from the simulation after reaching the arrestor ledge, to allow the ring to continue unperturbed. This was done by simply converting the rigid body volume fraction into void volume fraction.

The equation of state for the material is as follows. The cold-compression energy is again the standard form from \citet{RomenskiiEOS}, however the shear modulus and Gr\"{u}neisen function take the form outlined by \citet{Steinberg}, following \citet{BartonAnisotropicDamage, BartonLevelSetDamage}:
\begin{align}
 G_{(l)}(\rho_{(l)})      &= G_{0,(l)}+G_p p_c(\eta)\eta^{-1/3} \\
 \Gamma_{(l)}(\rho_{(l)}) &= \Gamma_{0,(l)} + \Gamma_{1,(l)}\eta^{-1.0/3.0}+ \Gamma_{2,(l)}\eta^{-\gamma_{(l)}} \ ,
\end{align}
where $\eta = \rho_{(l)}/\rho_{0,(l)}$ and $p_c = \rho_{(l)}^2\partial \mathscr{E}_c/ \partial \rho_{(l)}$. Parameters are given in Table \ref{tab:RomenskiiSteinburgParameters}. The Johnson-Cook plasticity model is employed, with parameters given in Table \ref{tab:DamagePlasticParameters}. 

A simplified damage model is used for this test, based on the model presented in \citet{BartonAnisotropicDamage} and \citet{BartonLevelSetDamage}. The model presented by \citet{BartonAnisotropicDamage} describes a fully anisotropic, tensorial damage model for use with the full deformation tensor representation, whereas this work employs an isotropic, scalar damage model for use with the unimodular stretch tensor, so some changes are required. This work proposes the following simple model for the damage evolution:
\begin{align}
 \dot{D} &= \dot{D_0}(1-D)\dot{\epsilon_p} \ ,
\end{align}
where $\dot{D_0}=0.15$ is a scalar multiplier. The material is considered fully damaged when $D=D_{\text{\scriptsize{crit}}}=0.1$. This corresponds to the damage dissipation potential of:
\begin{align}
 F^D = -\dot{D_0}Y \ .
\end{align}
This model is employed here so as to better match the results presented in \citet{BartonLevelSetDamage}, but as it is a simplified model, the \citet{Bonora} model is preferred in all other cases.

The test is run in three dimensions, in a domain spanning $x=[-30:30]$ mm, $y=[-30:30]$ mm, $z=[-0.8:0.8]$ mm. A base resolution of 300 $\times$ 300 $\times$ 8 cells is used, with 2 layers of AMR, each of refinement factor 2. The test is run for a time of 50 $\mu$s, using a CFL of 0.3. Again, a Lagrangian field damage perturbation is employed. Following \citet{BartonLevelSetDamage}, this work uses a Gaussian perturbation field, with a mean of 1 and a standard deviation of 0.167.

\begin{figure}
\centering
\includegraphics[width = 0.8\textwidth]{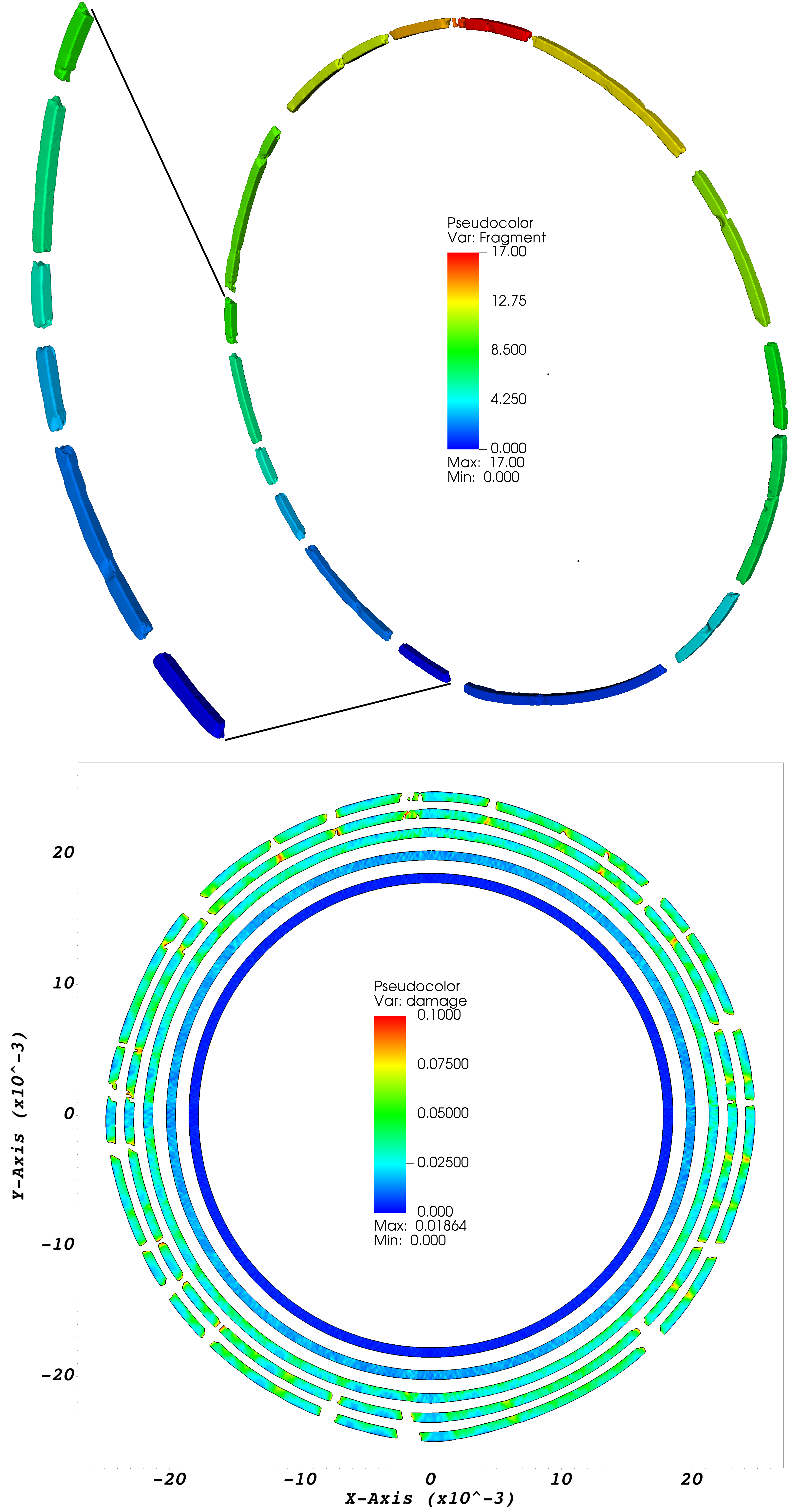}
\caption{The expanding ring test. The images depict (\textit{top}) the damage in the metal, shown at 0, 20, 30, 40 and 50 $\mu$s and (\textit{bottom}) the final fragments. The material is delineated from void by the 0.5 void volume fraction contour.}
\label{fig:ExpandingRing}
\end{figure}

\begin{figure}
\centering
\includegraphics[width = \textwidth]{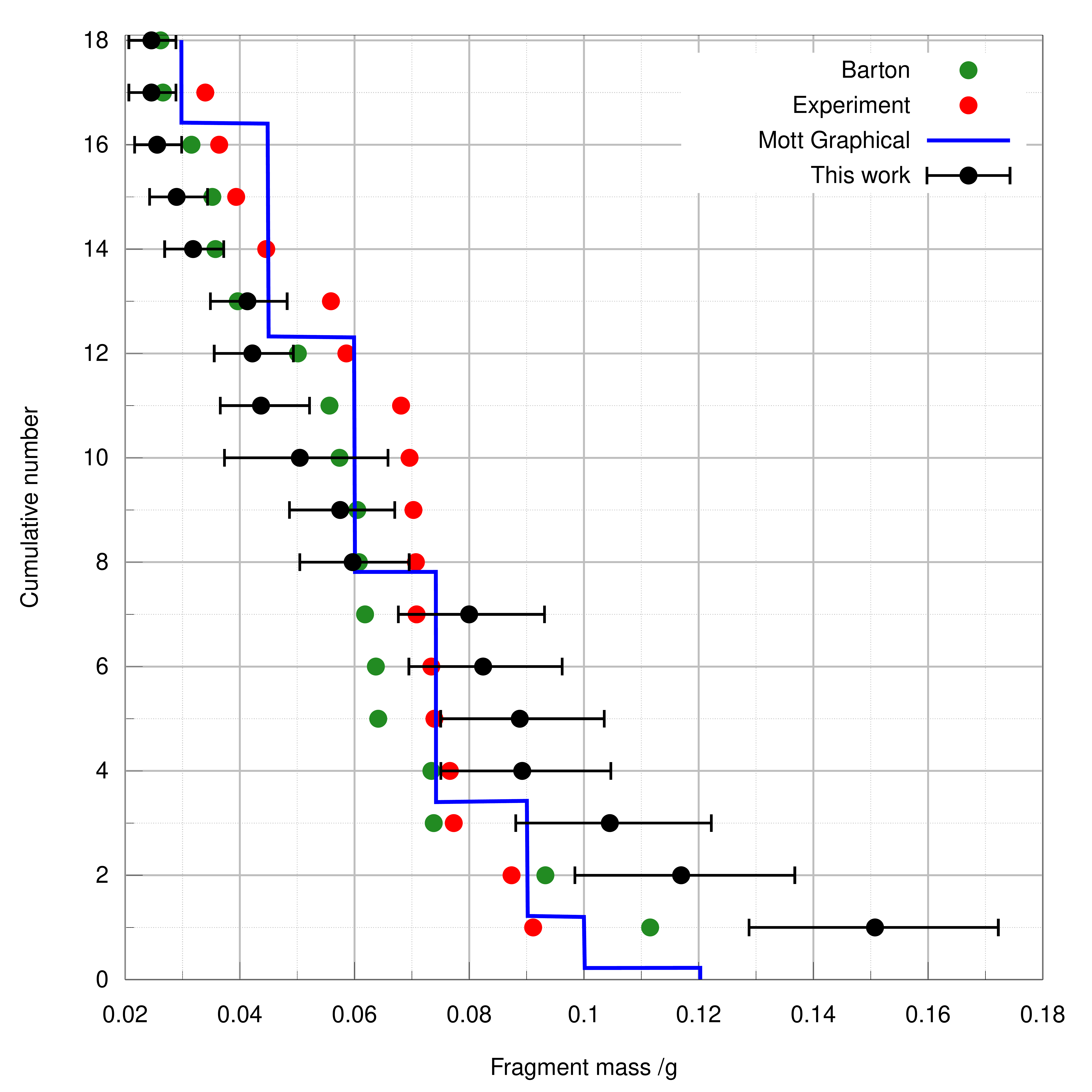}
\caption{The fragment mass distribution for the expanding ring test at 50 $\mu$s. The image shows the fragment mass against the cumulative fragment number for this work and the results presented in \citet{BartonLevelSetDamage}. The results agree well with both previous simulation and experiment.}
\label{fig:ExpandingRingFragments}
\end{figure}

The test is shown in Figure \ref{fig:ExpandingRing}, where the damage and fragments are plotted for various different times in the simulation. The measured fragment distribution is shown in Figure \ref{fig:ExpandingRingFragments}, and agrees well with the numerical and experimental results presented by \citet{BartonLevelSetDamage}. The results for this test represent a single simulation, whereas for full comparison to the statistical results, the model should be run multiple times. Nevertheless, the results are promising. Due to the diffuse interface nature of the method, there is also no single contour at which fragment boundaries can be demarcated. As such, results are plotted with errors showing several different values of the void volume fraction cut-off contour to give an indication of the range of the results. The lower bound is taken to be $\nu=0.25$, the middle is $\nu=0.5$ and the upper bound is $\nu=0.75$. Despite this range, the results still match well, thanks to the THINC interface sharpening keeping the interface diffusion to a minimum.

\subsection{Explosively driven shell}

\begin{figure}
\centering
\includegraphics[width = 0.9\textwidth]{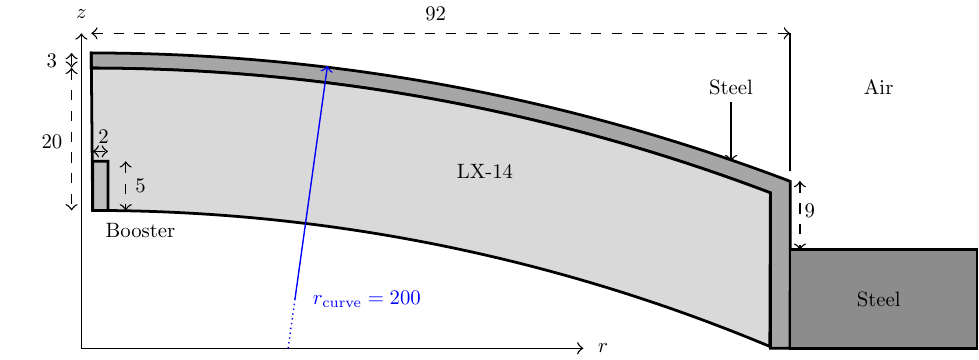}
\caption{The initial conditions for the explosive shell test. All lengths are given in mm.}
\label{fig:ExplosiveShellInitialConditions}
\centering
\includegraphics[width = 0.9\textwidth]{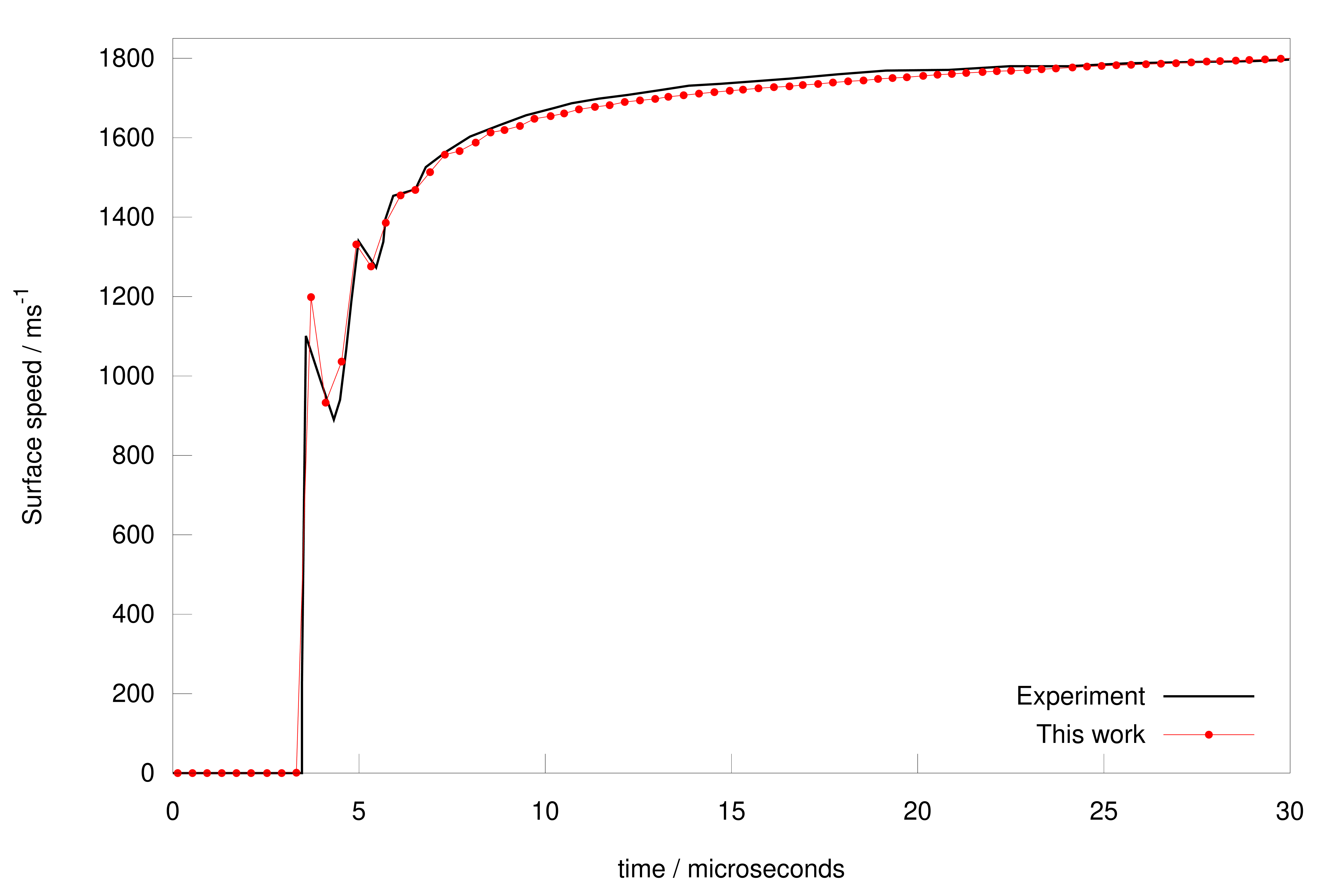}
\caption{A comparison of the measured and experimental velocity profile in the explosive shell test \citet{CampbellExplosiveShell}. The steel surface velocity was measured at $r = 46$ mm, and this profile was used to calibrate the strength of the explosive. The times reported by this work are also offset by $-2.3 \mu$s to match the time offset in the experiment.}
\label{fig:ExplosiveShellVelocityProfile}
\end{figure}

The final test considered in this work is an explosively driven shell test, taken from \citet{CampbellExplosiveShell}. In this test, a stainless steel shell is driven to fracture by high explosive. The test was designed to study fracture in biaxial tension at high strain rates, making it an excellent example for this work to consider. \citet{CampbellExplosiveShell} provide both experimental images and numerical data for comparison. There is also both experimental and numerical data from other studies \cite{LiExplosiveShell,OwenExplosiveShell} which use a variety of different numerical techniques. In many ways, this test represents the culmination of the methods developed in this work, as it features multi-material interaction, elastoplastic solids, high-explosive reactive fluid mixtures, material-void interaction, fracture, Lagrangian damage perturbation, adaptive mesh refinement, and three-dimensional geometry. The setup for the test is shown in Figure \ref{fig:ExplosiveShellInitialConditions}. The figure shows a radial slice of the domain, as the problem is axisymmetric.

The LX-14 high explosive is modelled using the method outlined in \citet{WallisMultiPhysics}. The LX-14 reactants and products both obey the JWL equation of state with the ignition and growth reaction rate law \cite{IgnitionAndGrowth}. This is a three-stage rate law, based on phenomenological experience of how detonations evolve in condensed phase explosives. The rate can be expressed as:
\begin{align}
\dot{\lambda} = I(1-F)^b\left(\frac{\rho}{\rho_0}-1-a\right)^x\text{H}(F_{ig}-F) + G_1(1-F)^cF^dp^y\text{H}(F_{G_1}-F) + G_2(1-F)^eF^gp^z\text{H}(F-F_{G_2}) \nonumber \ ,
\end{align}
where $F = 1-\lambda$ is the reacted fraction and $\text{H}$ is the Heaviside function. All other undefined parameters are material dependent constants. Parameters are given in Tables \ref{tab:DamageJWLParameters} and \ref{tab:DamageIandGParameters}. The stainless steel is modelled with the \citet{RomenskiiEOS} equation of state, using the Johnson-Cook plasticity model and \citet{Bonora} damage model, with parameters given in tables \ref{tab:DamageRomenskiiParameters}, \ref{tab:DamagePlasticParameters} and \ref{tab:DamageParameters}. Experimentally calibrated damage model parameters for the 304L stainless steel used in the original experiments of \citet{CampbellExplosiveShell} were not available, so the parameters were chosen so as to best match the experimental images provided by \citet{CampbellExplosiveShell}. The Johnson-Cook plasticity parameters for the steel were taken from \citet{Maurel-Pantel_304LSS_Plasticity}.

The test is run for a time of 65 $\mu$s, using a CFL of 0.3. The domain for the test spans $x=[-140:140]$ mm, $z=[-140:140]$ mm, $y=[-50:120]$ mm, with transmissive boundaries on all sides. Although the test is cylindrically symmetric, the test is run in three dimensions, without the use of geometric source terms. This allows for damage localisation and anisotropic fracture in the steel shell. Adaptive mesh refinement is particularly important in this test, due to the large domain size relative to the small steel shell. The test is run using 2 layers of adaptive mesh refinement, each of refinement factor 2, with a base mesh of $256\times256\times152$. Again, a Lagrangian damage perturbation field is used. This test uses a uniformly distributed field with a mean of 1 and a standard deviation of 0.2.

To ignite the explosive, the booster region in the centre of the shell has its pressure raised to 50 GPa, with all other materials being at atmospheric pressure.

A useful experimental comparison for this test is the surface velocity measured by \citet{CampbellExplosiveShell}. \citet{CampbellExplosiveShell} measure the surface velocity magnitude at a point on the steel surface midway between the pole and the edge. This work takes this to mean measuring the velocity magnitude at $r = 46$ mm. This velocity profile can then be used to calibrate the detonation energy, $Q$, of the explosive used in the test. The profile is shown in Figure \ref{fig:ExplosiveShellVelocityProfile}. This work finds that a value of $Q=6$ MJ kg$^{-1}$ best matches the experimental profile.

\begin{figure}
\centering
\includegraphics[height = 0.95\textheight]{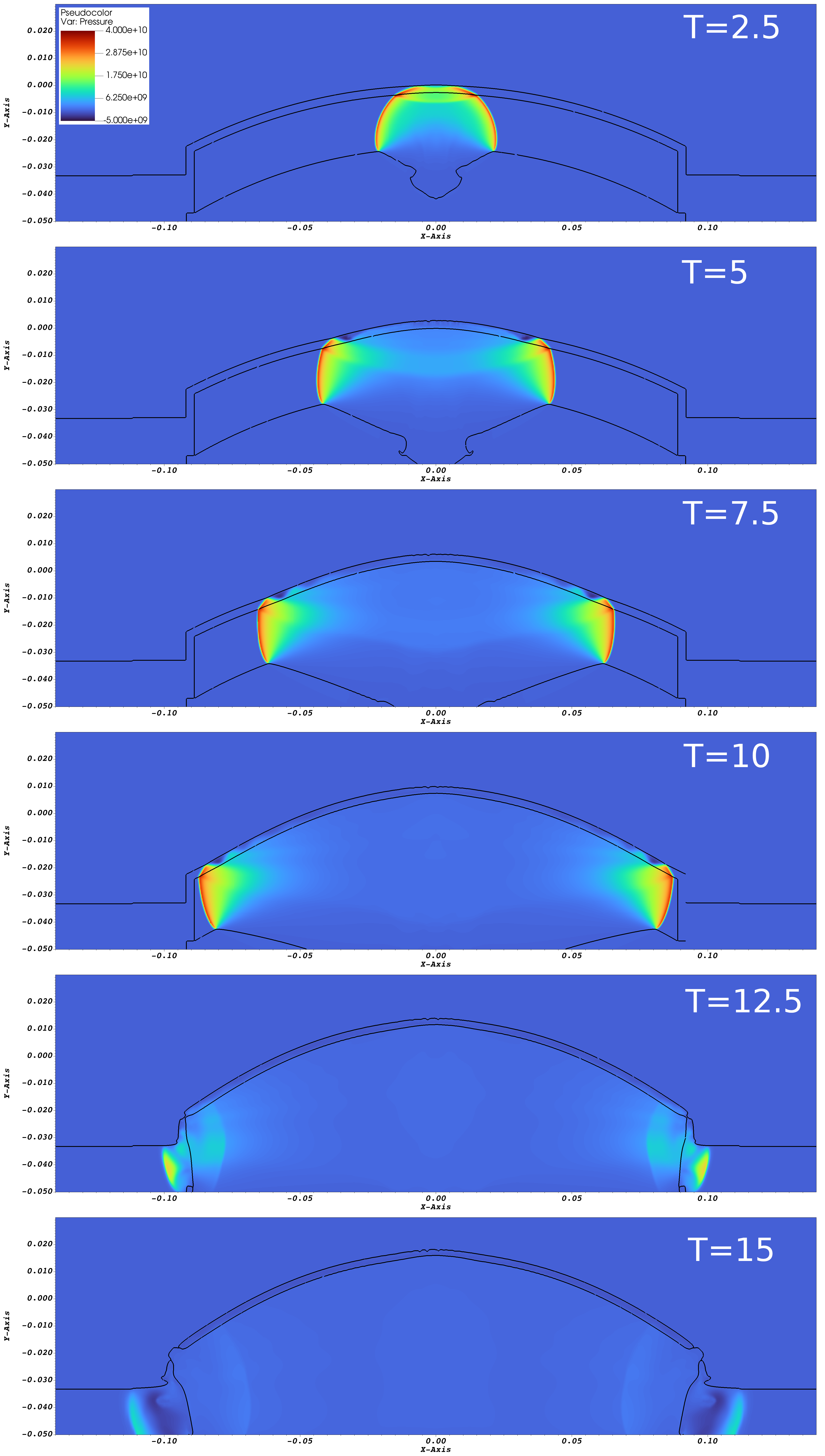}
\caption{Early times in the explosively driven shell test. The images depict the pressure in a slice through the centre of the domain. Times are shown in $\mu$s.}
\label{fig:ExplosiveShell_Slice}
\end{figure}

\begin{figure}
\centering
\includegraphics[width = \textwidth]{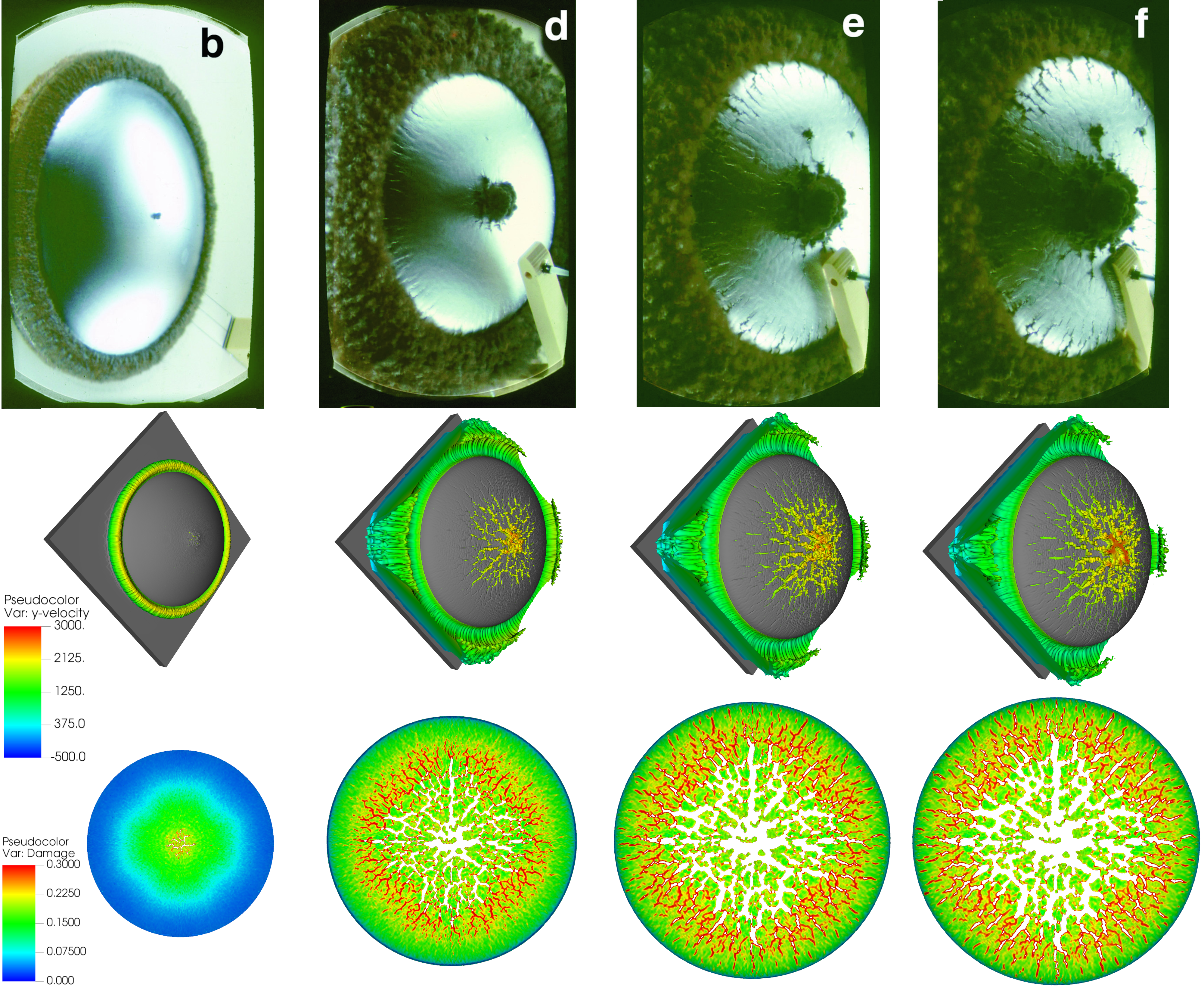}
\caption{The explosively driven shell test. (\textit{Top}) The experimental images from \citet{CampbellExplosiveShell}. (\textit{Middle}) The results of this work, at times corresponding to the experimental images: 22.5, 50, 60, 65 $\mu$s. The images depict the steel shell in grey and the z-velocity of the explosive products. (\textit{Bottom}) The damage in the steel shell. The experimental images are reprinted from Geoffrey H. Campbell, Gregory C. Archbold, Omar A. Hurricane, and Paul L. Miller , "Fragmentation in biaxial tension", Journal of Applied Physics 101, 033540 (2007) \cite{CampbellExplosiveShell}, with the permission of AIP Publishing.}
\label{fig:ExplosiveShell_Comparison}
\end{figure}

The test is shown in Figures \ref{fig:ExplosiveShell_Slice} and \ref{fig:ExplosiveShell_Comparison}. Figure \ref{fig:ExplosiveShell_Slice} shows the early evolution of the simulation, as the detonation wave propagates through the explosive, pushing the steel shell outwards. This image particularly demonstrates the explosive-solid coupling, as ringing can be seen in the steel shell as the detonation wave passes.

Figure \ref{fig:ExplosiveShell_Comparison} shows the later stages of the experiment. Here the steel shell begins to fracture, and the explosive products begin to seep through the cracks. These images correspond very well to the experimental images in \citet{CampbellExplosiveShell} as shown in the figure. The steel shell is shown in grey and the explosive products are coloured with their velocity along the axis of the shell. The fact that the explosive products can penetrate through the cracks in the steel shows the method is capable of handling damage and fracture in a multi-material context. Figure \ref{fig:ExplosiveShell_Comparison} also shows the damage in the later stages of the experiment. These images demonstrate the Lagrangian perturbation field working well to produce an anisotropic fracture pattern.

\section{Conclusions}

This work has outlined a three dimensional, multi-physics-compatible, diffuse interface method for fracture and fragmentation in realistic materials. This was achieved by extending the work of \citet{WallisFluxEnriched} to include realistic material inhomogeneity with the use of a scalar perturbation field that was updated using the semi-Lagrangian method of \citet{VitaliTransportDiffusion}. The method was tested on a variety of strenuous problems, and compared well to previous experiment and numerical simulation. The broad applicability of the underlying approach allowed the method to handle problems including rigid-body-driven fracture and explosively driven fracture with an explicitly resolved high explosive. This demonstrates the potential of the method for the simulation of a full experiment in one cohesive framework, rather than relying on co-simulation methods. Future work could include the addition of more realistic damage models, such as the anisotropic tensorial damage model presented in \citet{BartonAnisotropicDamage}, to further develop the model's capability to model realistic materials.

\appendix

\section{Material Parameters}

\begin{table*}[h]
\centering

\begin{tabular}[t]{|l|l|l|l|l|l|l|}
    \hline
    Material & $\rho_0$ \ kgm$^{-3}$ & $K_0$ \ GPa & $G_0$ \ GPa & $\alpha$ & $\beta$ & $\Gamma_0$ \\
    \hline
    CuBe           & 8370.0 & 131.3       & 53.6      & 1.0       & 3.0       & 2.0 \\
    Al-H32         & 2670.0 & 72.2        & 25.8      & 0.627354  & 2.28816   & 1.48389 \\
    Steel (304L SS)& 8000.0 & 158.3       & 73.1      & 0.569     & 2.437     & 1.84  \\
    \hline
\end{tabular}
\caption{The \citet{RomenskiiEOS} equation of state material parameters. *A constant shear modulus variant was used for this material.}
\label{tab:DamageRomenskiiParameters}

\begin{tabular}[t]{|l|l|l|l|l|l|l|l|l|l|}
    \hline
    Material & $\rho_0$ \ kgm$^{-3}$ & $K_0$ \ GPa & $G_0$ \ GPa & $G_p$ & $\alpha$ & $\Gamma_0$ & $\Gamma_1$ & $\Gamma_2$ & $\gamma$ \\
    \hline
    U-Nb     & 17411.0 & 111.27 & 26.34 & 1.1226 & 1.105 & 0.5 & 1.056 & 1.068 & 3.843 \\
    \hline
\end{tabular}
\caption{The \citet{RomenskiiEOS} equation of state material parameters with the \citet{Steinberg} form for the shear modulus and Gr\"{u}neisen function.}
\label{tab:RomenskiiSteinburgParameters}

\begin{tabular}[t]{|l|l|l|l|l|l|l|l|}
    \hline
    Material          & $c_1$  GPa & $c_2$  GPa & $c_3$ & $n$  & $m$ & $T_{\text{melt}}$  & $C_V$ J kg$^{-1}$ K$^{-1}$ \\
    \hline
    CuBe              & 1.041 & 0.0   & 0.025 & 0.31  & - & - & - \\ %1.09 & 1286.0  \\
    Al-H32            & 0.275 & 0.114 & 0.002 & 0.42  & - & - & -\\
    Steel (304L SS)   & 0.253 & 0.685 & 0.097 & 0.313 & 2.044 & 1689.0 & 468.6\\
    U-Nb              & 0.780 & 0.253 & 0.012 & 0.22  & 1.0   & 1710.0 & 115.0\\
    \hline
\end{tabular}
\caption{Johnson Cook plasticity material parameters.}
\label{tab:DamagePlasticParameters}

\begin{tabular}[t]{|l|l|l|l|l|l|}
    \hline
    Material & $D_\text{crit}$ & $\epsilon_{p,\text{thresh}}$ & $\epsilon_{p,\text{crit}}$ & $\alpha$ & $\nu$ \\
    \hline
    CuBe              & 0.85 & 0.08 & 0.16 & 0.631 & 0.324 \\
    Steel (304L SS) (Expanding Disk) & 0.3  & 0.002  & 0.5  & 0.5   & 0.3 \\
    Steel (304L SS) (Explosive Shell) & 0.3  & 0.5  & 1.2  & 0.5   & 0.3 \\
    \hline 
\end{tabular}
\caption{\citet{Bonora} damage model parameters.}
\label{tab:DamageParameters}

\begin{tabular}{|c|c|c|c|c|c|c|c|}
\hline
Material & $\rho_0$ & ${\cal A}$ / $10^{11}$ Pa & ${\cal B}$ / $10^{11}$ Pa & ${\cal R}_1$ & ${\cal R}_2$ & $\Gamma$ & $C^V$ / J kg$^{-1}$K$^{-1}$ \\
\hline
Reactant & 1850   & 7320.0 & -0.052654 & 14.1 & 1.41 & 0.8938 & 1461.6\\
Product  & -      & 16.689 & 0.5969 & 5.9 & 2.1 & 0.45 & 540.5 \\
\hline
\end{tabular} 
\caption{The JWL equation of state parameters for LX-14, taken from \citet{LX14JWL}.}~\\[0.2cm]
\label{tab:DamageJWLParameters}

\begin{tabular}{|c|c|c|c|c|c|c|c|c|c|c|c|}
\hline
$a$ & $b$ & $c$ & $d$ & $e$ & $g$ & $x$ & $y$ & $z$ & $F_{ig}$ & $F_{G_1}$ & $F_{G_2}$ \\
\hline
0.0819 & 0.667 & 0.667 & 0.45 & 0.667 & 0.5 & 4 & 2 & 4 & 0.02 & 1.0 & 0.02 \\
\hline
\end{tabular}~\\[0.1cm]
\begin{tabular}{|c|c|c|c|}
\hline
I /s$^{-1}$ & G$_1$ / ($10^{11}$ Pa)$^{-y}$ s$^{-1}$   & G$_2$ / ($10^{11}$ Pa)$^{-z}$ s$^{-1}$ & $Q$ MJ kg$^{-1}$ \\
\hline
$2\times 10^{10}$& 170 $\times 10^{6}$ & 2 $\times 10^{10}$ & 6.0\\
\hline
\end{tabular}
\caption{The dimensional ignition and growth reaction rate parameters for LX-14 from \citet{LX14JWL}.}
\label{tab:DamageIandGParameters}
\end{table*}

\clearpage

\bibliography{references}

\end{document}